\newif\iflinenums
\linenumsfalse
\def\docopts{preprint}
\def\docclass{aastex}
\def\tabletype{deluxetable}

\ifdefined\apj
  \linenumsfalse
  \def\docopts{manuscript}
  \def\docclass{aastex}
  \def\tabletype{deluxetable}
\fi

\ifdefined\arxiv
  \linenumsfalse
  \def\docclass{aastex}  
  \def\tabletype{deluxetable} 
\fi

\documentclass[\docopts]{\docclass}

\usepackage{graphicx}
\usepackage[dvipsnames,svgnames]{xcolor} 
\usepackage{natbib} 
\usepackage{hyperref}
\usepackage{amsmath}
\usepackage{multirow}
\usepackage{accents}
\hypersetup{    
  colorlinks      = true,
  linkcolor       = {black},
  linkbordercolor = {white},
  citecolor       = {black},
  citebordercolor = {white},
  urlcolor        = {blue},
  urlbordercolor  = {white},
}
\usepackage[figure,figure*]{hypcap}

\usepackage{soul} 
\usepackage{amsmath}
\usepackage{xspace}
\usepackage{lineno}
\usepackage{verbatim}

\usepackage{aas_macros}

\graphicspath{{./}{./figs/}}

\newcommand{\eg}{e.g.\xspace}
\newcommand{\etc}{etc.\xspace}
\newcommand{\etal}{et al.\xspace}

\mathchardef\mhyphen="2D
\newcommand{\vect}[1]{\boldsymbol{#1}}
\newcommand{\roughly}{\ensuremath{ {\sim}\,} }

\newlength{\dhatheight}

\newcommand{\code}[1]{\texttt{#1}\xspace}
\newcommand{\dd}{\ensuremath{\rm d}}

\newcommand{\unit}[1]{\ensuremath{\mathrm{\,#1}}\xspace}
\newcommand{\Gyr}{\unit{Gyr}}

\newcommand{\pc}{\unit{pc}}
\newcommand{\kpc}{\unit{kpc}}
\newcommand{\second}{\unit{s}}

\newcommand{\Msolar}{\ensuremath{M_\odot}}

\newcommand{\Lsun}{\ensuremath{L_\odot}}

\newcommand{\magn}{\unit{mag}}

\newcommand{\secref}[1]{Section~\ref{sec:#1}}

\newcommand{\tabref}[1]{Table~\ref{tab:#1}}

\newcommand{\figref}[1]{Figure~\ref{fig:#1}}

\newcommand{\eqnref}[1]{Equation~\eqref{eqn:#1}}

\newcommand{\nobjs}{{eight}\xspace}

\newcommand{\retII}{{DES\,J0335.6\allowbreak$-$5403}\xspace}
\newcommand{\eriII}{{DES\,J0344.3\allowbreak$-$4331}\xspace}
\newcommand{\tucII}{{DES\,J2251.2\allowbreak$-$5836}\xspace}
\newcommand{\horI}{{DES\,J0255.4\allowbreak$-$5406}\xspace}
\newcommand{\picI}{{DES\,J0443.8\allowbreak$-$5017}\xspace}
\newcommand{\indI}{{DES\,J2108.8\allowbreak$-$5109}\xspace}
\newcommand{\eriIII}{{DES\,J0222.7\allowbreak$-$5217}\xspace}
\newcommand{\pheII}{{DES\,J2339.9\allowbreak$-$5424}\xspace}

\newcommand{\SExtractor}{\code{SExtractor}}
\newcommand{\HEALPix}{\code{HEALPix}}

\newcommand{\PARSEC}{\code{PARSEC}}
\newcommand{\mangle}{\code{mangle}}
\newcommand{\emcee}{\code{emcee}}
\newcommand{\UGALI}{\code{UGALI}}

\newcommand{\modulus}{\ensuremath{M - m}\xspace}
\newcommand{\spreadmodel}{\ensuremath{spread\_model}\xspace}
\newcommand{\magauto}{\ensuremath{mag\_auto}\xspace}
\newcommand{\magpsf}{\ensuremath{mag\_psf}\xspace}

\newcommand{\ra}{{\ensuremath{\alpha_{2000}}}\xspace}
\newcommand{\dec}{{\ensuremath{\delta_{2000}}}\xspace}
\newcommand{\age}{{\ensuremath{\tau}}\xspace}
\newcommand{\metal}{{\ensuremath{Z}}\xspace}

\newcommand{\TS}{\ensuremath{\mathrm{TS}}\xspace}

\newcommand{\like}{\ensuremath{\mathcal{L}}\xspace} 
\newcommand{\loglike}{\ensuremath{\log\like}\xspace}
\newcommand{\given}{\ensuremath{ \,|\, }\xspace}

\newcommand{\data}{ \ensuremath{ \mathcal{D} }\xspace } 

\newcommand{\params}{\ensuremath{\vect{\theta}}\xspace}

\newcommand{\pdf}{PDF\xspace}

\newcommand{\uspatial}{\ensuremath{u_s}}
\newcommand{\ucolor}{\ensuremath{u_c}}

\providecommand\physrep{\ref@jnl{Phys.~Rep.}}%
\providecommand\apjs{\ref@jnl{ApJS}}%
\providecommand{\jcap}{\ref@jnl{JCAP}}%

\begin{document} 
\iflinenums
  \linenumbers
\fi

\title{Eight New Milky Way Companions Discovered in First-Year Dark Energy Survey Data} 

\author{
K.~Bechtol\altaffilmark{1,\dagger},
A.~Drlica-Wagner\altaffilmark{2,\dagger},
E.~Balbinot\altaffilmark{3,4},
A.~Pieres\altaffilmark{5,4},
J.~D.~Simon\altaffilmark{6},
B.~Yanny\altaffilmark{2},
B.~Santiago\altaffilmark{5,4},
R.~H.~Wechsler\altaffilmark{7,8,11},
J.~Frieman\altaffilmark{2,1},
A.~R.~Walker\altaffilmark{9},
P.~Williams\altaffilmark{1},
E.~Rozo\altaffilmark{10,11},
E.~S.~Rykoff\altaffilmark{11},
A.~Queiroz\altaffilmark{5,4},
E.~Luque\altaffilmark{5,4},
A.~Benoit-L{\'e}vy\altaffilmark{12},
D.~Tucker\altaffilmark{2},
I.~Sevilla\altaffilmark{13,14},
R.~A.~Gruendl\altaffilmark{15,13},
L.~N.~da Costa\altaffilmark{16,4},
A.~Fausti Neto\altaffilmark{4},
M.~A.~G.~Maia\altaffilmark{4,16},
T.~Abbott\altaffilmark{9},
S. ~Allam\altaffilmark{17,2},
R.~Armstrong\altaffilmark{18},
A.~H.~Bauer\altaffilmark{19},
G.~M.~Bernstein\altaffilmark{18},
R.~A.~Bernstein\altaffilmark{6},
E.~Bertin\altaffilmark{20,21},
D.~Brooks\altaffilmark{12},
E.~Buckley-Geer\altaffilmark{2},
D.~L.~Burke\altaffilmark{11},
A.~Carnero Rosell\altaffilmark{4,16},
F.~J.~Castander\altaffilmark{19},
R.~Covarrubias\altaffilmark{15},
C.~B.~D'Andrea\altaffilmark{22},
D.~L.~DePoy\altaffilmark{23},
S.~Desai\altaffilmark{24,25},
H.~T.~Diehl\altaffilmark{2},
T.~F.~Eifler\altaffilmark{26,18},
J.~Estrada\altaffilmark{2},
A.~E.~Evrard\altaffilmark{27},
E.~Fernandez\altaffilmark{28,39},
D.~A.~Finley\altaffilmark{2},
B.~Flaugher\altaffilmark{2},
E.~Gaztanaga\altaffilmark{19},
D.~Gerdes\altaffilmark{27},
L.~Girardi\altaffilmark{16},
M.~Gladders\altaffilmark{29,1},
D.~Gruen\altaffilmark{30,31},
G.~Gutierrez\altaffilmark{2},
J.~Hao\altaffilmark{2},
K.~Honscheid\altaffilmark{32,33},
B.~Jain\altaffilmark{18},
D.~James\altaffilmark{9},
S.~Kent\altaffilmark{2},
R.~Kron\altaffilmark{1},
K.~Kuehn\altaffilmark{34,35},
N.~Kuropatkin\altaffilmark{2},
O.~Lahav\altaffilmark{12},
T.~S.~Li\altaffilmark{23},
H.~Lin\altaffilmark{2},
M.~Makler\altaffilmark{36},
M.~March\altaffilmark{18},
J.~Marshall\altaffilmark{23},
P.~Martini\altaffilmark{33,37},
K.~W.~Merritt\altaffilmark{2},
C.~Miller\altaffilmark{27,38},
R.~Miquel\altaffilmark{28,39},
J.~Mohr\altaffilmark{24},
E.~Neilsen\altaffilmark{2},
R.~Nichol\altaffilmark{22},
B.~Nord\altaffilmark{2},
R.~Ogando\altaffilmark{4,16},
J.~Peoples\altaffilmark{2},
D.~Petravick\altaffilmark{15},
A.~A.~Plazas\altaffilmark{40,26},
A.~K.~Romer\altaffilmark{41},
A.~Roodman\altaffilmark{7,11},
M.~Sako\altaffilmark{18},
E.~Sanchez\altaffilmark{14},
V.~Scarpine\altaffilmark{2},
M.~Schubnell\altaffilmark{27},
R.~C.~Smith\altaffilmark{9},
M.~Soares-Santos\altaffilmark{2},
F.~Sobreira\altaffilmark{2,4},
E.~Suchyta\altaffilmark{32,33},
M.~E.~C.~Swanson\altaffilmark{15},
G.~Tarle\altaffilmark{27},
J.~Thaler\altaffilmark{42},
D.~Thomas\altaffilmark{22},
W.~Wester\altaffilmark{2},
J.~Zuntz\altaffilmark{43}
\\ \vspace{0.2cm} (The DES Collaboration) \\
}
\altaffiltext{\textdagger}{Co-first authors: \href{mailto:bechtol@kicp.uchicago.edu}{bechtol@kicp.uchicago.edu}, \href{mailto:kadrlica@fnal.gov}{kadrlica@fnal.gov} }
\altaffiltext{1}{Kavli Institute for Cosmological Physics, University of Chicago, Chicago, IL 60637, USA}
\altaffiltext{2}{Fermi National Accelerator Laboratory, P. O. Box 500, Batavia, IL 60510, USA}
\altaffiltext{3}{Department of Physics, University of Surrey, Guildford GU2 7XH, UK}
\altaffiltext{4}{Laborat\'orio Interinstitucional de e-Astronomia - LIneA, Rua Gal. Jos\'e Cristino 77, Rio de Janeiro, RJ - 20921-400, Brazil}
\altaffiltext{5}{Instituto de F\'\i sica, UFRGS, Caixa Postal 15051, Porto Alegre, RS - 91501-970, Brazil}
\altaffiltext{6}{Carnegie Observatories, 813 Santa Barbara St., Pasadena, CA 91101, USA}
\altaffiltext{7}{Kavli Institute for Particle Astrophysics \& Cosmology, P. O. Box 2450, Stanford University, Stanford, CA 94305, USA}
\altaffiltext{8}{Department of Physics, Stanford University, 382 Via Pueblo Mall, Stanford, CA 94305, USA}
\altaffiltext{9}{Cerro Tololo Inter-American Observatory, National Optical Astronomy Observatory, Casilla 603, La Serena, Chile}
\altaffiltext{10}{University of Arizona, Department of Physics, 1118 E. Fourth St., Tucson, AZ 85721, U.S.A.}
\altaffiltext{11}{SLAC National Accelerator Laboratory, Menlo Park, CA 94025, USA}
\altaffiltext{12}{Department of Physics \& Astronomy, University College London, Gower Street, London, WC1E 6BT, UK}
\altaffiltext{13}{Department of Astronomy, University of Illinois,1002 W. Green Street, Urbana, IL 61801, USA}
\altaffiltext{14}{Centro de Investigaciones Energ\'eticas, Medioambientales y Tecnol\'ogicas (CIEMAT), Madrid, Spain}
\altaffiltext{15}{National Center for Supercomputing Applications, 1205 West Clark St., Urbana, IL 61801, USA}
\altaffiltext{16}{Observat\'orio Nacional, Rua Gal. Jos\'e Cristino 77, Rio de Janeiro, RJ - 20921-400, Brazil}
\altaffiltext{17}{Space Telescope Science Institute, 3700 San Martin Drive, Baltimore, MD  21218, USA}
\altaffiltext{18}{Department of Physics and Astronomy, University of Pennsylvania, Philadelphia, PA 19104, USA}
\altaffiltext{19}{Institut de Ci\`encies de l'Espai, IEEC-CSIC, Campus UAB, Facultat de Ci\`encies, Torre C5 par-2, 08193 Bellaterra, Barcelona, Spain}
\altaffiltext{20}{Sorbonne Universit\'es, UPMC Univ Paris 06, UMR 7095, Institut d'Astrophysique de Paris, F-75014, Paris, France}
\altaffiltext{21}{CNRS, UMR 7095, Institut d'Astrophysique de Paris, F-75014, Paris, France}
\altaffiltext{22}{Institute of Cosmology \& Gravitation, University of Portsmouth, Portsmouth, PO1 3FX, UK}
\altaffiltext{23}{George P. and Cynthia Woods Mitchell Institute for Fundamental Physics and Astronomy, and Department of Physics and Astronomy, Texas A\&M University, College Station, TX 77843,  USA}
\altaffiltext{24}{Department of Physics, Ludwig-Maximilians-Universitat, Scheinerstr.\ 1, 81679 Munich, Germany}
\altaffiltext{25}{Excellence Cluster Universe, Boltzmannstr.\ 2, 85748 Garching, Germany}
\altaffiltext{26}{Jet Propulsion Laboratory, California Institute of Technology, 4800 Oak Grove Dr., Pasadena, CA 91109, USA}
\altaffiltext{27}{Department of Physics, University of Michigan, Ann Arbor, MI 48109, USA}
\altaffiltext{28}{Institut de F\'{\i}sica d'Altes Energies, Universitat Aut\`onoma de Barcelona, E-08193 Bellaterra, Barcelona, Spain}
\altaffiltext{29}{Department of Astronomy and Astrophysics, University of Chicago, Chicago IL 60637, USA}
\altaffiltext{30}{Max Planck Institute for Extraterrestrial Physics, Giessenbachstrasse, 85748 Garching, Germany}
\altaffiltext{31}{University Observatory Munich, Scheinerstrasse 1, 81679 Munich, Germany}
\altaffiltext{32}{Department of Physics, The Ohio State University, Columbus, OH 43210, USA}
\altaffiltext{33}{Center for Cosmology and Astro-Particle Physics, The Ohio State University, Columbus, OH 43210, USA}
\altaffiltext{34}{Australian Astronomical Observatory, North Ryde, NSW 2113, Australia}
\altaffiltext{35}{Argonne National Laboratory, 9700 S. Cass Avenue, Lemont IL 60639 USA}
\altaffiltext{36}{ICRA, Centro Brasileiro de Pesquisas F\'isicas, Rua Dr. Xavier Sigaud 150, CEP 22290-180, Rio de Janeiro, RJ, Brazil}
\altaffiltext{37}{Department of Astronomy, The Ohio State University, Columbus, OH 43210, USA}
\altaffiltext{38}{Department of Astronomy, University of Michigan, Ann Arbor, MI, 48109, USA}
\altaffiltext{39}{Instituci\'o Catalana de Recerca i Estudis Avan\c{c}ats, E-08010, Barcelona, Spain}
\altaffiltext{40}{Brookhaven National Laboratory, Bldg 510, Upton, NY 11973, USA}
\altaffiltext{41}{Astronomy Centre, University of Sussex, Falmer, Brighton, BN1 9QH, UK}
\altaffiltext{42}{Department of Physics, University of Illinois, 1110 W. Green St., Urbana, IL 61801, USA}
\altaffiltext{43}{Jodrell Bank Center for Astrophysics, School of Physics and Astronomy, University of Manchester, Oxford Road, Manchester, M13 9PL, UK}


\begin{abstract}

We report the discovery of \nobjs new Milky Way companions in $\roughly 1{,}800 \deg^2$ of optical imaging data collected during the first year of the Dark Energy Survey (DES).
Each system is identified as a statistically significant over-density of individual stars consistent with the expected isochrone and luminosity function of an old and metal-poor stellar population.
The objects span a wide range of absolute magnitudes ($M_V$ from $-2.2 \magn$ to $-7.4 \magn$), physical sizes ($10\pc$ to $170\pc$), and heliocentric distances ($30\kpc$ to $330\kpc$).
Based on the low surface brightnesses, large physical sizes, and/or large Galactocentric distances of these objects, several are likely to be new ultra-faint satellite galaxies of the Milky Way and/or Magellanic Clouds.
We introduce a likelihood-based algorithm to search for and characterize stellar over-densities, as well as identify stars with high satellite membership probabilities.
We also present completeness estimates for detecting ultra-faint galaxies of varying luminosities, sizes, and heliocentric distances in the first-year DES data. 
\end{abstract}

\maketitle 

\section{Introduction}
\label{sec:intro}

Milky Way satellite galaxies provide a unique opportunity to study the low-luminosity threshold of galaxy formation and to better connect the baryonic component of galaxies with the dark matter halos in which they reside.
Prior to the Sloan Digital Sky Survey (SDSS), the faintest known galaxies had luminosities of $\roughly 10^{5} \Lsun$, and it was clear that the population of twelve ``classical'' Milky Way satellites was orders of magnitude smaller than would be naively expected in the cold dark matter paradigm~\citep{1999ApJ...522...82K,1999ApJ...524L..19M}.
Over the past decade, systematic searches of wide-field SDSS imaging have revealed fifteen additional arcminute-scale, resolved stellar over-densities 
\citep{2005AJ....129.2692W,2005ApJ...626L..85W,2006ApJ...650L..41Z,2006ApJ...643L.103Z,2006ApJ...647L.111B,2007ApJ...654..897B,2008ApJ...686L..83B,2009MNRAS.397.1748B,2010ApJ...712L.103B,2006ApJ...645L..37G,2009ApJ...693.1118G,2006ApJ...653L..29S,2007ApJ...656L..13I,2007ApJ...662L..83W}
that have been either photometrically classified or spectroscopically confirmed as gravitationally-bound ``ultra-faint'' galaxies
\citep{2005ApJ...630L.141K,2006ApJ...650L..51M,2007MNRAS.380..281M,2007ApJ...670..313S,2009ApJ...692.1464G,2009ApJ...690..453K,2009MNRAS.397.1748B,2009ApJ...694L.144W,2009ApJ...702L...9C,2009A&A...506.1147A,2011AJ....142..128W,2011ApJ...736..146K,2011ApJ...733...46S,2013ApJ...770...16K}.
These ultra-faint galaxies are the smallest, least luminous, least chemically enriched, and most dark matter dominated galaxies in the known Universe.

Since all known ultra-faint Milky Way satellite galaxies were discovered in SDSS, the census of these objects is almost certainly incomplete due to the partial sky coverage ($\roughly 14{,}000 \deg^{2}$) and photometric magnitude limit (95\% complete to $r \sim 22 \magn$) of that survey.
While only 27 Milky Way satellite galaxies are currently known, extrapolations of the luminosity function suggest that hundreds of luminous Milky Way satellites remain to be found in current and near-future wide-field optical imaging surveys~\citep{2008ApJ...688..277T,2014ApJ...795L..13H,PhysRevD.91.063515}. 

The Dark Energy Survey (DES) is in the process of imaging $5{,}000 \deg^{2}$ of the southern Galactic cap in five photometric bands \citep{Abbott:2005bi,2014SPIE.9149E..0VD}.
The deep photometry of DES ($r \sim 24$ mag) will enable the detection of the faintest known satellite galaxies out to $\roughly 120\kpc$ (compared to the SDSS limit of $\roughly 50\kpc$), and more luminous satellite galaxies out to the Milky Way virial radius~\citep{2011AJ....141..185R}.
We have completed an initial search of the first year of DES data and report here on the \nobjs most significant dwarf galaxy candidates discovered therein (\tabref{detection}).
Since the physical nature of these candidates cannot be definitively determined with photometry alone, we refer to them by their discovery coordinates.
If these candidates are later confirmed to be Local Group galaxies, they should be renamed after the constellation in which they reside: \retII (Reticulum~II), \eriII (Eridanus~II), \tucII (Tucana~II), \horI (Horologium~I), \indI (Indus~I), \picI (Pictor~I), \pheII (Phoenix~II), and \eriIII (Eridanus~III).
If any are instead globular clusters, they would be known as DES 1 through N.
After the completion of this work, we learned that \indI was previously identified by \citet{2015ApJ...803...63K} in data from the Stromlo Milky Way Satellite Survey and designated as a likely star cluster, Kim~2.

\begin{figure*}
  \includegraphics[width=1.\textwidth]{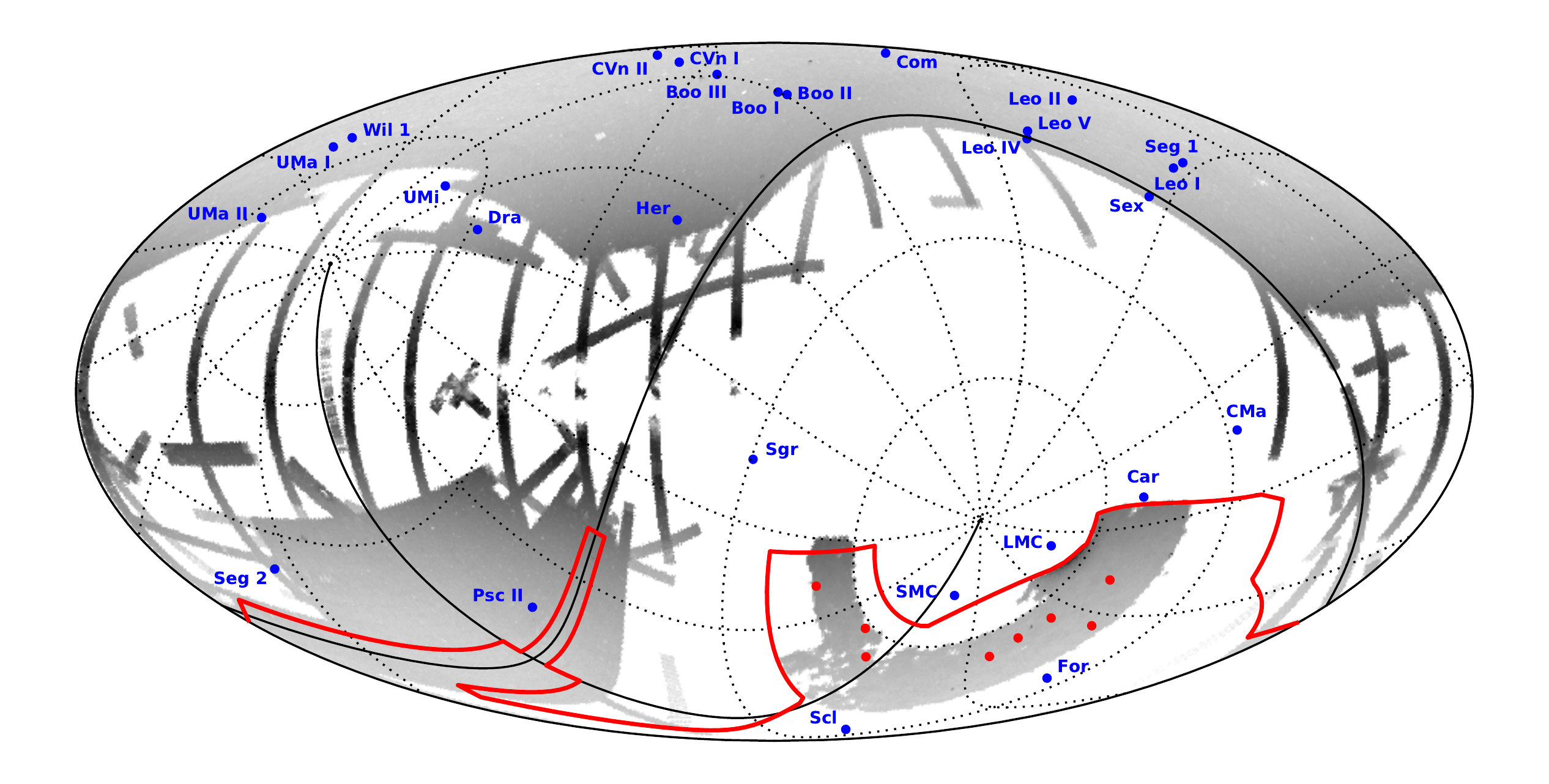}
  \caption{Locations of 27 known Milky Way satellite galaxies \citep[blue;][]{McConnachie12} and \nobjs DES dwarf galaxy candidates (red) in Galactic coordinates (Mollweide projection). 
  The coordinate grid shows the equatorial coordinate system with solid lines for the equator and zero meridian. 
  The gray scale indicates the logarithmic density of stars with $r < 22$ from SDSS and DES.
  The large contiguous region in the northern equatorial hemisphere shows the coverage of SDSS \citep{2014ApJS..211...17A}.
  The full DES footprint is outlined in red, and is now partially filled in by a region of $\roughly 1{,}600\deg^2$ near to the Magellanic Clouds and a region of $\roughly 200 \deg^2$ overlapping with the SDSS Stripe 82 field along the celestial equator.
  Both fields were observed during the first year of DES and that compose the Y1A1 data set.
  }
\label{fig:skymap}
\end{figure*}

\section{Data Set}
\label{sec:data}

DES is a wide-field optical imaging survey in the $grizY$ bands performed with the Dark Energy Camera \citep[DECam; ][]{2010SPIE.7735E..0DF,DECamPaper2015,flaugher_2015_decam}.
The DECam focal plane comprises 74 CCDs: 62 2k$\times$4k CCDs dedicated to science imaging and 12 2k$\times$2k CCDs for guiding, focus, and alignment. 
DECam is installed at the prime focus of the 4-meter Blanco telescope at Cerro Tololo Inter-American Observatory.
In this configuration, DECam has a hexagonal 2.2-degree-wide field-of-view and a central pixel scale of 0.263 arcseconds. 
The full DES survey is scheduled for 525 nights distributed over five years.
Here, we consider data collected between 15 August 2013 and 9 February 2014 during the first year of DES.

The first internal annual release of DES data (Y1A1) consists of $\roughly 12{,}000$ science exposures processed by the DES data management (DESDM) infrastructure (Gruendl et al., in preparation).\footnote{\url{http://data.darkenergysurvey.org/aux/releasenotes/DESDMrelease.html}}
Most of the Y1A1 footprint is covered by 2 to 4 overlapping exposures, or ``tilings'', in each filter.
Single exposures in a tiling are $90 \second$ in $griz$ and $45 \second$ in $Y$.
Here, we rely on the $g$- and $r$-band images for photometry, and use the $i$-band for star-galaxy separation. 

The DESDM image processing pipeline consists of image detrending, astrometric calibration, nightly photometric calibration, global calibration, image coaddition, and object catalog creation, as recently summarized in \citet{2015MNRAS.449.1129B}.
We refer to \citet{Sevilla:2011}, \citet{2012ApJ...757...83D}, and \citet{2012SPIE.8451E..0DM} for a more detailed description of the DES single-epoch and coadd image processing.
The \SExtractor toolkit is used to create object catalogs from the processed and coadded images \citep{2011ASPC..442..435B, 1996A&AS..117..393B}. 
The Y1A1 data release contains a catalog of $\roughly 131$ million unique objects detected in the coadd imaging which are distributed over $1{,}800 \deg^2$.
This area includes $\roughly 200 \deg^2$ overlapping with the Stripe 82 region of SDSS, as well as a contiguous region of $\roughly 1{,}600\deg^2$ overlapping the South Pole Telescope (SPT) footprint~\citep{Carlstrom:2009um}. 
The DES imaging in the SPT region is unprecedented in depth. 
\figref{skymap} shows the coverage of Y1A1 in Galactic coordinates.


We selected stars from the Y1A1 coadd object catalog based on the \spreadmodel quantity output by \SExtractor~\citep{2012ApJ...757...83D}.
To avoid issues arising from fitting the point-spread function (PSF) across variable-depth coadded images, we utilized the weighted-average ($wavg$) of the \spreadmodel measurements from the single-epoch exposures.
Our stellar sample consists of well-measured objects with $|wavg\_spread\_model\_i| \allowbreak < \allowbreak 0.003$, $flags\_\{g,r,i\} < 4$, and $magerr\_auto\_\{g,r,i\} < 1$.
We also removed objects for which the \magpsf and \magauto measurements differ by more than $0.5 \magn$ because this is indicative of poor object characterization. 

We estimated the stellar completeness on a statistical basis over the full Y1A1 footprint by creating a test sample of high stellar purity using a color-based selection of $r - i > 1.7$.
We then applied the morphology-based star selection criteria above that uses imaging in only a single band to evaluate the stellar completeness for the test sample.
This approach is unbiased since the two star selection criteria are orthogonal, and benefits from a large statistical sample that is representative of the full Y1A1 data set.
The stellar completeness was found to be ${>}\,90\%$ to $g \sim 22 \magn$ and falls to $\roughly 50\%$ by $g \sim 23 \magn$.
We validated this completeness estimate with matched spectroscopic data in the overlap region between Y1A1 and SDSS Stripe 82.
Based on studies with the \code{DAOPHOT}\footnote{\url{http://www.star.bris.ac.uk/~mbt/daophot/}} software package optimized for photometry in crowded stellar fields, we do not expect this stellar completeness to be reduced in the vicinity of DES satellite galaxy candidates relative to the Y1A1 footprint at large.

For point-like objects and a well-estimated PSF, the \SExtractor \magpsf variables are expected to give the best measurement of stellar fluxes.
However, due to the aforementioned difficulties with PSF estimation on deep coadded images, we chose instead to use the \magauto measurements.
The \magauto measurements are found to give a less biased estimate of flux when compared to a stellar calibration sample from Pan-STARRS~\citep{2012ApJ...756..158S}.
Measured magnitudes are extinction corrected using the $E(B-V)$ dust maps from~\citet{1998ApJ...500..525S}.
The relative calibration uncertainties are estimated via stellar-locus regression \citep{Kelly:2014} and are found to be $\roughly 2\%$ across the survey footprint.
Uncertaities in the offsets between the DES photometric system and the AB system are estimated to be $\roughly 1\%$.

\section{Search Methods}
\label{sec:methods}

\begin{\tabletype}{l cccc cccc cc}
\tablecolumns{13}
\tablewidth{0pt}
\tabletypesize{\tiny}
\tablecaption{Detection of new satellite galaxy candidates in DES Y1A1\label{tab:detection}}
\tablehead{
Name & $\alpha_{2000}$ & $\delta_{2000}$  & $m-M$ & Map Sig    & TS Scan & TS Fit & $r_h$      & $\epsilon$ & $\phi$   & $\Sigma p_i$ \\
     &  $(\deg)$       & $(\deg)$         &       & $(\sigma)$ &         &        & $(\deg)$   &            & $(\deg)$ & }
\startdata
DES\,J0335.6-5403 (Ret\,II) & 53.92  & -54.05 & 17.5 & 24.6 & 1466 & 1713 & $0.10^{+0.01}_{-0.01}$    & $0.6^{+0.1}_{-0.2}$    & $72^{+7}_{-7}$   & 338.1 \\
DES\,J0344.3-4331 (Eri\,II) & 56.09  & -43.53 & 22.6 & 23.0 & 322  & 512  & $0.03^{+0.01}_{-0.01}$    & $0.19^{+0.16}_{-0.16}$ & $90^{+30}_{-30}$ & 96.9 \\
DES\,J2251.2-5836 (Tuc\,II) & 343.06 & -58.57 & 18.8 & 6.4  & 129  & 167  & $0.12^{+0.03}_{-0.03}$    & -- & -- & 114.9 \\
DES\,J0255.4-5406 (Hor\,I)  & 43.87  & -54.11 & 19.7 & 8.2  & 55   & 81   & $0.04^{+0.05}_{-0.02}$    & -- & -- & 30.6 \\
DES\,J2108.8-5109 (Ind\,I)  & 317.20 & -51.16 & 19.2 & 5.5  & --   & 75   & $0.010^{+0.002}_{-0.002}$ & -- & -- & 26.6 \\
DES\,J0443.8-5017 (Pic\,I)  & 70.95  & -50.28 & 20.5 & 7.1  & --   & 63   & $0.02^{+0.07}_{-0.01}$    & -- & -- & 19.1 \\
DES\,J2339.9-5424 (Phe\,II) & 354.99 & -54.41 & 19.9 & 5.1  & --   & 61   & $0.02^{+0.01}_{-0.01}$    & -- & -- & 19.4 \\
DES\,J0222.7-5217 (Eri\,III)& 35.69  & -52.28 & 19.9 & 5.4  & --   & 57   & $0.007^{+0.005}_{-0.003}$ & -- & -- & 8.9 \\
\tableline

\enddata
{\footnotesize
\tablecomments{Best-fit parameters from the maximum-likelihood fit assuming the composite isochrone described in \secref{likelihood}.
 Uncertainties are calculated from the the highest density interval containing 90\% of the posterior distribution.
``Map Sig'' refers to detection significance of the candidate from the stellar density map search method (\secref{visual}). 
``TS Scan'' refers to the significance (Equation \ref{eqn:ts}) from the likelihood scan using a Plummer model spatial kernel with half-light radius $r_h = 0\fdg1$ (\secref{likelihood}). 
``TS Fit'' denotes the significance of the likelihood method using the set of best-fit parameters.
Ellipticities and position angles are not quoted for lower significance candidates where they are not well constrained by the data. 
For objects with significant ellipticity, the half-light radius is measured along the elliptical semi-major axis.
$\Sigma p_i$ is the estimated number of satellite member stars with $g < 23$ in the stellar catalog.}
}
\end{\tabletype}

Ultra-faint galaxies are discovered as arcminute-scale over-densities of individually resolved stars.
The Y1A1 stellar object catalog is of such quality and depth that numerous stellar over-densities are immediately apparent from a visual scan.
Several of these over-densities are not associated with any known star cluster, globular cluster, or satellite galaxy.
To formalize the process of identifying new candidate satellite galaxies, we applied both (1) a simple spatial binning algorithm to facilitate inspection of the stellar density field, and (2) a matched-filter maximum-likelihood technique. 
These complementary approaches validated one another and the resultant list of candidates was vetted by both methods.

\subsection{Stellar Density Maps}
\label{sec:visual}

Several independent searches of the stellar density field were conducted.
One approach involved direct visual inspection of coadded images.
Other searches used binned stellar density maps constructed from the coadd object catalogs. 
As an example, we detail below how one of these maps was built and analyzed.


We began by spatially binning the stellar catalog into equal-area pixels using the \HEALPix scheme~\citep{2005ApJ...622..759G}.%
\footnote{\url{http://healpix.sourceforge.net}}
We considered \HEALPix pixel sizes of $\roughly0\fdg06$ ($nside = 1024$) and $\roughly0\fdg11$ ($nside = 512$) to optimize sensitivity to satellites possessing different angular sizes.
Since the stellar density is greatly enhanced in regions of the Y1A1 footprint near the LMC and Galactic plane, we further grouped the stars into larger regions of $\sim13 \deg^2$ ($nside = 16$) to estimate the local field density of stars.
We corrected the effective solid angle of each pixel using the survey coverage, as estimated by \mangle as part of DESDM processing~\citep{2008MNRAS.387.1391S}.\footnote{\url{http://space.mit.edu/~molly/mangle/}}
Several conspicuous stellar over-densities were immediately apparent after this simple spatial binning procedure.

We increase our sensitivity to ultra-faint satellite galaxies by focusing our search on regions of color-magnitude space populated by old, low-metallicity stellar populations \citep{2008ApJ...686..279K,2009AJ....137..450W}.
As a template, we used a \PARSEC isochrone corresponding to a stellar population of age $12\Gyr$ and metallicity $\metal=0.0002$~\citep{2012MNRAS.427..127B}.
Sensitivity to satellites at varying distances was enhanced by considering 20 logarithmically spaced steps in heliocentric distance ranging from $20\kpc$ to $400\kpc$ (distance moduli $16.5 < \modulus < 23.0$).
For each step in distance, all stars within $0.2\magn$ of the isochrone in magnitude-magnitude space were retained while those outside the isochrone template were discarded.
We then created a significance map for each $\sim13 \deg^2$ region by computing the Poisson likelihood of finding the observed number of stars in each map pixel given a background level characterized by the local field density. 

\subsection{Matched-filter Maximum-likelihood Method}
\label{sec:likelihood}

The simple approach described above is computationally efficient and easily generalizable.
However, a more sensitive search can be performed by simultaneously modeling the spatial and photometric distributions of stars and incorporating detailed characteristics of the survey (variable depth, photometric uncertainty, \etc).
One way to incorporate this information is through a maximum-likelihood analysis~\citep{1925PCPS...22..700F,edwards1972likelihood}.
Likelihood-based analyses have found broad applicability in studies of Milky Way satellites \citep[\eg,][]{2002MNRAS.332...91D,2008ApJ...684.1075M}.
Here we extend the maximum-likelihood approach to a wide-area search for Milky Way satellites.
Similar strategies have been applied to create catalogs of galaxy clusters over wide-field optical surveys \citep[\eg,][]{2014ApJ...785..104R}.

Our maximum-likelihood search begins by assuming that the stellar catalog in a small patch of sky represents a Poisson realization of (1) a field contribution including Milky Way foreground stars, mis-classified background galaxies, and imaging artifacts, and (2) a putative satellite galaxy.
The unbinned Poisson log-likelihood function is given by
\begin{equation}
\label{eqn:loglike}
   \loglike = -f \lambda + \sum_i \left ( 1 - p_i \right ),
\end{equation}
\noindent where $i$ indexes the objects in the stellar sample. 
The value $p_i$ can be interpreted as the probability that star $i$ is a member of the satellite, and is computed as
\begin{equation}
\label{eqn:membership}
  p_i \equiv \frac{\lambda u_i}{\lambda u_i + b_i}.
\end{equation}
\noindent Here, $u$ represents the signal probability density function (\pdf) for the satellite galaxy and is normalized to unity over the spatial and magnitude domain, $\mathcal{S}$; specifically, $\int_{\rm all} u\,\dd{\mathcal{S}} = 1$. 
The corresponding background density function for the field population is denoted by $b$. 

\begin{figure*}
  \includegraphics[width=1.\textwidth]{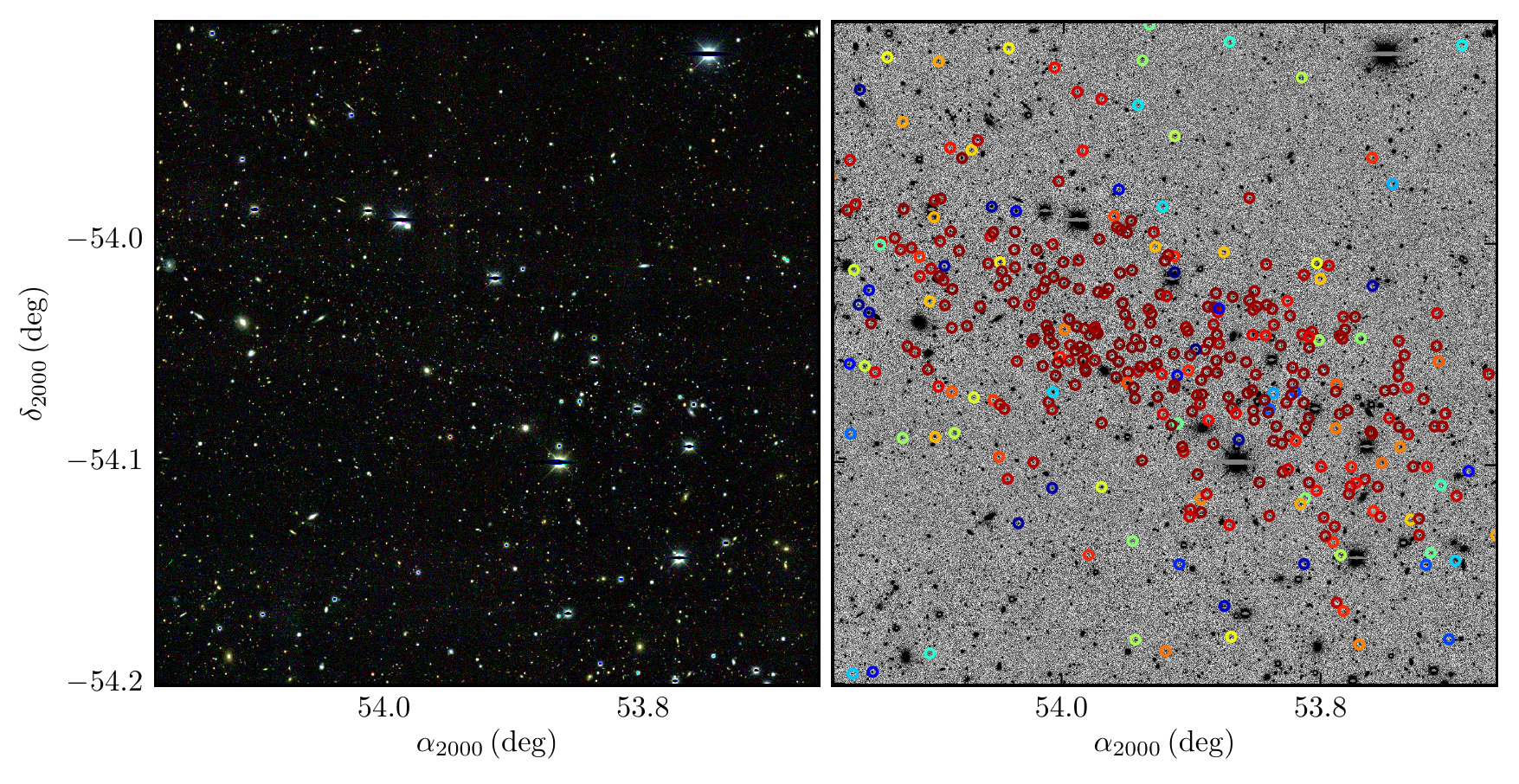}
  \caption{\textit{Left}: False color $gri$ coadd image of the $0\fdg3 \times 0\fdg3$ region centered on \retII.
  	\textit{Right}: Stars in the same field of view with membership probability $p_i > 0.01$ are marked with colored circles.  
	In this color map, red signifies high-confidence association with \retII and blue indicates lower membership probability.
	The membership probabilities have been evaluated using \eqnref{membership} for the best-fit model parameters listed in \tabref{detection}.
  }
\label{fig:coadd_image}
\end{figure*}

We define the richness, $\lambda$, to be a normalization parameter representing the total number of satellite member stars with mass ${>}0.1 \Msolar$.
In \eqnref{loglike}, $f \equiv \int_{\rm obs} u\, \dd{\mathcal{S}}$ represents the fraction of satellite member stars that are within the \textit{observable} spatial and magnitude domain of the survey, and $f \lambda$ denotes the expected number of observable satellite member stars.
\footnote{\code{Mangle} maps of the survey coverage are used in the calculation of the observable fraction at each position in the sky.}
Maximizing the likelihood with respect to the richness implies $f \lambda = \sum_i p_i$.
This condition makes clear that the satellite membership probability for each star in the catalog is a natural product of the maximum-likelihood approach.
These membership probabilities can be used to prioritize targeting when planning spectroscopic follow-up observations.
\figref{coadd_image} highlights the use of membership probabilities to visualize a low-surface-brightness satellite galaxy candidate.

To characterize a candidate satellite galaxy, we explore the likelihood of the data, \data, as a function of a set of input model parameters, \params.
The signal \pdf is assumed to be separable into two independent components,
\begin{equation}
u(\data_i \given \params) = \uspatial(\data_{s,i} \given \params_s)
\times \ucolor(\data_{c,i} \given \params_c).
\end{equation}
\noindent The first component, \uspatial, depends only on the spatial properties, while the second component, \ucolor, depends only on the distribution in color-magnitude space.

We modeled the spatial distribution of satellite member stars with an elliptical Plummer profile~\citep{1911MNRAS..71..460P}, following the elliptical coordinate prescription of \citet{2008ApJ...684.1075M}. 
The Plummer profile is sufficient to describe the spatial distribution of stars in known ultra-faint galaxies~\citep{2012ApJ...745..127M}.
The spatial data for catalog object $i$ consist of spatial coordinates, $\data_{s,i} = \{\alpha_i, \delta_i\}$, while the parameters of our elliptical Plummer profile are the centroid coordinates, half-light radius, ellipticity, and position angle, $\params_{s} = \{\alpha_0, \delta_0, r_h, \epsilon, \phi\}$.

We modeled the color-magnitude component of the signal \pdf with a set of representative isochrones for old, metal-poor stellar populations, specifically by taking a grid of isochrones from \citet{2012MNRAS.427..127B} spanning $0.0001 < \metal < 0.001$ and $1 \Gyr < \age < 13.5 \Gyr$.
Our spectral data for star $i$ consist of the magnitude and magnitude error in each of two filters, $\data_{c,i} = \{ g_i, \sigma_{g,i}, r_i, \sigma_{r,i}\}$, while the model parameters are composed of the distance modulus, age, and metallicity describing the isochrone, $\params_{c} = \{ \modulus, \age, \metal \}$.
To calculate the spectral signal \pdf, we weight the isochrone by a \citet{2001ApJ...554.1274C} initial mass function (IMF) and densely sample in magnitude-magnitude space.
We then convolve the photometric measurement \pdf of each star with the \pdf of the weighted isochrone.
The resulting distribution represents the predicted probability of finding a star at a given position in magnitude-magnitude space given a model of the stellar system. 

The background density function of the field population is empirically determined from a circular annulus surrounding each satellite candidate ($ 0\fdg5 < r < 2\fdg0$).
The inner radius of the annulus is chosen to be sufficiently large that the stellar population of the candidate satellite does not bias the estimate of the field population.
Stellar objects in the background annulus are binned in color-magnitude space using a cloud-in-cells algorithm and are weighted by the inverse solid angle of the annulus.
The effective solid angle of the annulus is corrected to account for regions that are masked or fall below our imposed magnitude limit of $g < 23 \magn$. 
The resulting two-dimensional histogram for the field population provides the number density of stellar objects as a function of observed color and magnitude ($\deg^{-2} \magn^{-2}$).
This empirical determination of the background density incorporates contamination from unresolved galaxies and imaging artifacts.

The likelihood formalism above was applied to the Y1A1 data set via an automated analysis pipeline.\footnote{The Ultra-faint Galaxy Likelihood (\UGALI) code; detailed methodology and performance to be presented elsewhere.}
For the search phase of the algorithm, we used a radially symmetric Plummer model with half-light radius $r_h = 0\fdg1$ as the spatial kernel, and a composite isochrone model consisting of four isochrones bracketing a range of ages, $\age = \{ 12 \Gyr, 13.5 \Gyr \}$, and metallicities, $\metal = \{ 0.0001,0.0002 \}$, to bound a range of possible stellar populations.
We then tested for a putative satellite galaxy at each location on a three-dimensional grid of sky position (0.7 arcmin resolution; $nside = 4096$) and distance modulus ($16 <\modulus< 24$; $16\kpc$ to $630\kpc$).

The statistical significance at each grid point can be expressed as a Test Statistic (\TS) based on the likelihood ratio between a hypothesis that includes a satellite galaxy versus a field-only hypothesis:
\begin{equation}
\label{eqn:ts}
\TS = 2 \left[\loglike(\lambda = \hat{\lambda}) - \loglike(\lambda = 0)\right].
\end{equation}
\noindent Here, $\hat{\lambda}$ is the value of the stellar richness that maximizes the likelihood. 
In the asymptotic limit, the null-hypothesis distribution of the \TS will follow a $\chi^2 / 2$ distribution with one bounded degree of freedom~\citep{chernoff1954likelihood}.
We have verified that the output distribution of our implementation agrees well with the theoretical expectation by testing on simulations of the stellar field.
In this case, the \textit{local} statistical significance of a given stellar over-density, expressed in Gaussian standard deviations, is approximately the square root of the \TS.

\section{Candidate Selection and Characterization}
\label{sec:results}

The two search methods described in \secref{methods} each produce significance maps of the full Y1A1 footprint, where peaks in these maps represent the three-dimensional seed positions (\ra, \dec, \modulus) of possible satellite galaxies.
Seeds were selected from the union of the search methods.
Statistical significance thresholds were set at ${>}\,5 \sigma$ for the stellar density map method and $\TS > 45$ for the matched-filter maximum-likelihood method, yielding $\roughly50$ seeds.
Most of these were discarded as being attributed to steep gradients in the stellar density field, numerical effects near the survey boundaries, imaging artifacts, and large nearby galaxies resolved into multiple closely spaced catalog objects.
For this reason, we did not pursue investigation at lower significance thresholds.

The resulting seed list was compared against catalogs of known star clusters \citep[][2010 edition]{2013A&A...558A..53K,Harris96} and Milky Way satellite galaxies \citep{McConnachie12} as well as catalogs of other astrophysical objects that can produce false positives, such as large nearby galaxies~\citep{1973ugcg.book.....N,2004yCat.7239....0H} or galaxy clusters \citep{2014ApJ...785..104R}.  
Associated seeds include the Reticulum globular cluster, the Phoenix dwarf galaxy, AM\,1, NGC\,1261, NGC\,1291, NGC\,1553, NGC\,1851, NGC\,7089, NGC\,7424, ESO\,121-SC\,003, and ESO\,201-SC\,010.

We explored the multi-dimensional parameter space for each unassociated seed using the \emcee module for Markov Chain Monte Carlo~\citep[MCMC; ][]{2013PASP..125..306F},\footnote{\emcee v2.1.0: \url{http://dan.iel.fm/emcee/current/}} and the likelihood function described in \secref{likelihood} with flat priors on each of the input parameters.
For each seed, we ran an MCMC chain with 100 walkers that each make 1000 steps including a burn-in stage of 50 steps. 
This is sufficient to sample the region of parameter space near the maximum-likelihood estimate.
Only seeds with well-constrained posterior distributions enter our candidate list of new Milky Way companions.

\tabref{detection} presents the \nobjs most significant stellar over-densities in the Y1A1 data set consistent with being previously unknown dwarf galaxies.
When comparing the significances obtained with the map-based and likelihood scan algorithms, it is worth noting that the two methods were applied assuming different size scales for the target satellites, and that kernel assumed for the likelihood scan ($r_h = 0\fdg1$) is larger than the majority of candidates listed in \tabref{detection}.
After fitting the spatial parameters of the candidates, all are detected with high significance using the likelihood-based method.
The dependence of detection efficiency on assumed kernel size is discussed is \secref{completeness}.

The physical characteristics of these objects, as determined by the follow-up MCMC likelihood analysis, are summarized in \tabref{properties}.
The best-fit values and uncertainties are determined from the peak of the posterior distribution and the 90\% highest posterior density interval \citep{box1973bayesian}.
A significant correlation was observed between the age of the stellar isochrone and the heliocentric distance --- a degeneracy that may be expected given the evolution of the main sequence turnoff.
For some DES candidates, the posterior distribution for the distance is multi-modal.
The distance estimates provided in \tabref{properties} indicate the peaks in the posterior distribution.

To compare with previously known Milky Way satellite galaxies, we convert from DES $g$- and $r$-band magnitudes to visual magnitudes using
\begin{equation}
\begin{aligned}
g_{\rm DES} &= g_{\rm SDSS} - 0.104(g_{\rm SDSS} - r_{\rm SDSS}) + 0.01 \magn \\
r_{\rm DES} &= r_{\rm SDSS} - 0.102(g_{\rm SDSS} - r_{\rm SDSS}) + 0.02 \magn \\
\Rightarrow V &= g_{\rm DES} - 0.487(g_{\rm DES} - r_{\rm DES}) - 0.025 \magn.
\end{aligned}
\end{equation}
\noindent This transform from DES $g$ and $r$ magnitudes to $V$-band magnitudes was derived using an SDSS stellar calibration sample and the equations from \citet{2005AJ....130..873J}.
The absolute magnitude of each satellite is calculated using the sampling formalism of \citet{2008ApJ...684.1075M}.
For bright satellites, this formalism yields a very similar estimate to the integration of the stellar luminosity function for the best-fit model.
However, for fainter satellites, the uncertainty in the total magnitude can be dominated by shot noise arising from sparse sampling of the stellar population.
In this case, the additional association of a single bright star can have a strong influence on the measured magnitude of a satellite.
Similarly, the evolution of individual member stars can substantially change the total luminosity.
To quantify the impact of shot noise on the derived luminosity estimates, we use a representative isochrone weighted by a Chabrier IMF to simulate an ensemble of satellites with similar characteristics to the observed candidates.
The quoted uncertainty on the luminosity reflects the expected shot noise from stars in the magnitude range visible to DES, $17 \magn < g < 23 \magn$.

The angular and physical half-light radii listed in \tabref{detection} and \tabref{properties} are both given as two-dimensional quantities.
The deprojected (three-dimensional) half-light radius is a factor $\sim1.3$ larger than the projected half-light radius for a variety of common density profiles \citep{2010MNRAS.406.1220W}.
For objects with measured ellipticity, we report the half-light radius measured along the semi-major axis.

As illustrated in \figref{skymap}, the DES candidates are distributed throughout the Y1A1 footprint and occupy a portion of the celestial sphere in the direction of the Magellanic Clouds where no ultra-faint galaxies were previously known.
The DES candidates are widely distributed in heliocentric distance from $\roughly 30\kpc$ (\retII) to ${>}\,300 \kpc$ (\eriII).

\section{Discussion}

\begin{figure*}
  \includegraphics[width=1.\textwidth]{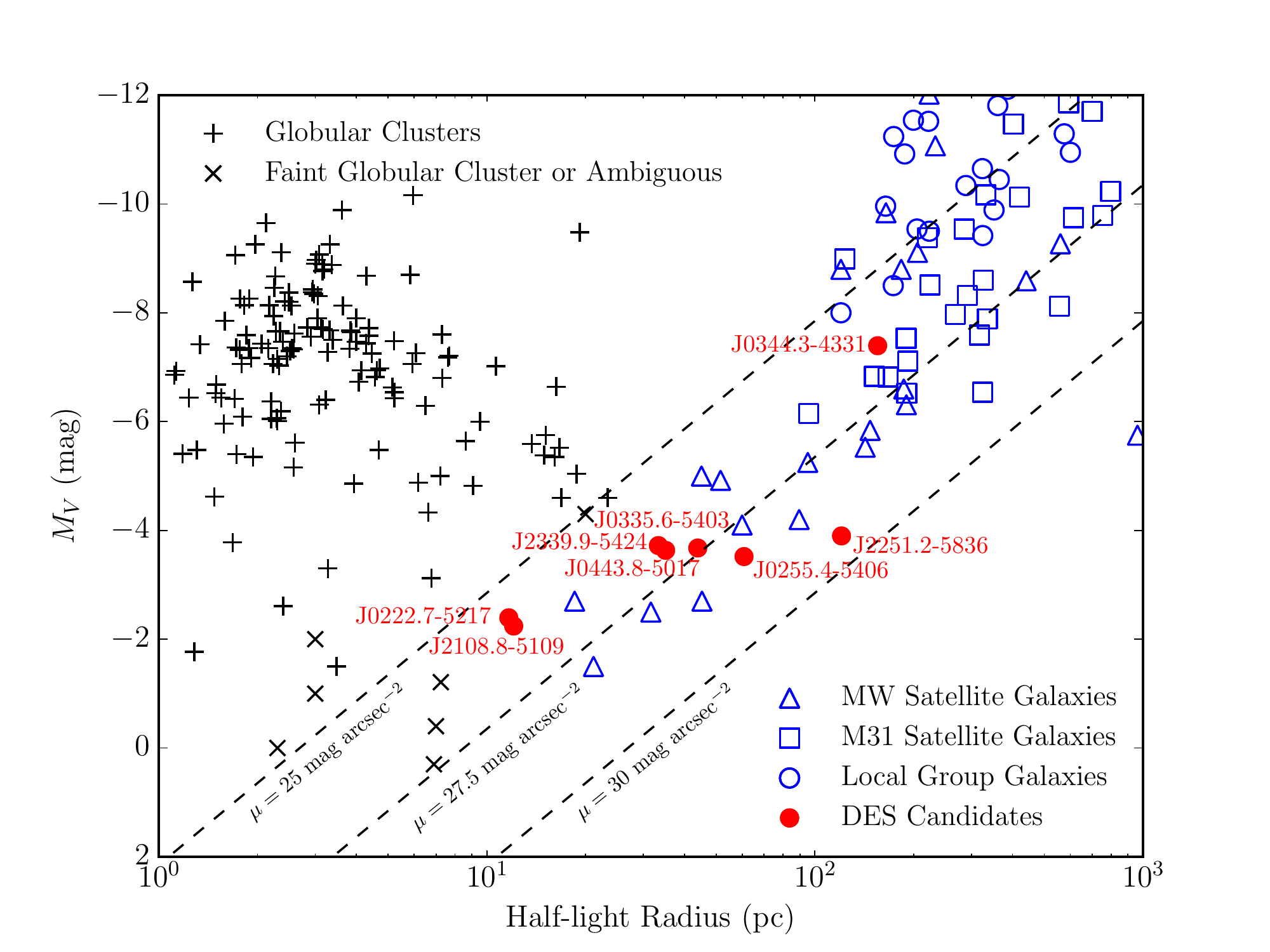}
  \caption{Local Group galaxies \citep{McConnachie12} and globular clusters \citep[][2010 edition]{Harris96} occupy distinct regions in the plane of physical half-light radius (geometric mean of the major and minor axes) and absolute luminosity. 
  The majority of DES satellite candidates (red dots) are more consistent with the locus of Local Group galaxies (empty blue shapes) than with the population of Galactic globular clusters (black crosses).
  Several of the faintest globular clusters and systems of ambiguous classification are indicated with $\times$ marks: Koposov~1 and Koposov~2 \citep{2007ApJ...669..337K,2014AJ....148...19P}, Segue~3 \citep{2010ApJ...712L.103B,2011AJ....142...88F,2013MNRAS.433.1966O}, Mu\~{n}oz~1 \citep{2012ApJ...753L..15M}, Balbinot~1 \citep{2013ApJ...767..101B}, PSO~J174.0675-10.8774 / Crater~I \citep{2014ApJ...786L...3L,2014MNRAS.441.2124B}, and Kim~1 \citep{2015ApJ...799...73K}.
  Dashed lines indicate contours of constant surface brightness at $\mu = \{25, 27.5, 30\} \magn \unit{arcsec}^{-2}$.
}
\label{fig:size_luminosity}
\end{figure*}

Galaxies are distinguished from star clusters by having a dynamical mass that is substantially larger than the mass inferred from the luminous stellar population and/or a significant dispersion in metallicities indicative of multiple generations of star formation and a deep enough gravitational potential to retain supernova ejecta~\citep{2012AJ....144...76W}.
While devoted spectroscopic follow up observations are necessary to unambiguously classify these objects, the properties given in \tabref{properties} already provide strong clues as to which candidates are most likely to be galaxies.
First, the large physical sizes of most of these objects are more consistent with the locus occupied by known satellite galaxies of the Local Group than with globular clusters of the Milky Way, as shown in \figref{size_luminosity}.
All of the DES candidates are of comparable surface brightness to the ultra-faint galaxies detected in SDSS~\citep{McConnachie12}.
The two most compact systems, \indI and \eriIII, fall in between the known ultra-faint galaxies and the faintest Milky Way star clusters, e.g., Koposov~1 and Koposov~2 \citep{2007ApJ...669..337K,2014AJ....148...19P}, Segue~3 \citep{2010ApJ...712L.103B,2011AJ....142...88F,2013MNRAS.433.1966O}, Mu\~{n}oz~1 \citep{2012ApJ...753L..15M}, Balbinot~1 \citep{2013ApJ...767..101B},  and Kim~1 \citep{2015ApJ...799...73K}.
For the most significant DES candidates, it is possible to estimate the ellipticity.
Whereas globular clusters tend to have ellipticity $\lesssim0.2$ \citep{2008AJ....135.1731V,2008ApJ...684.1075M}, the best measured candidate, \retII, has an ellipticity $\sim0.6$, which is more consistent with the population of known ultra-faint galaxies.

Further insight can be gained by fitting isochrones to the observed stellar distribution in color-magnitude space.
Two independent maximum-likelihood implementations confirm that the DES candidates are generally consistent with old ($\tau \gtrsim 10$~Gyr) and metal-poor stellar populations ($\metal \sim 0.0002$). 
The first of these analyses is the pipeline described in \secref{likelihood} used in a mode that varies age and metallicity in addition to spatial parameters and distance modulus in a simultaneous fit.
The second color-magnitude fitting procedure adopts a similar likelihood formalism, but fits the spatial and photometric distributions of the stars in two separate phases.
Instead of assuming an initial mass function, the second method weights the stars according to their proximity to the best-fit centroid location, and then evaluates the consistency between each star and a given isochrone in color-magnitude space given the photometric uncertainty for that star.
The second method is more robust to complications that might arise from stellar incompleteness and/or imperfect modeling of the initial mass function.
Age estimates and metallicity upper limits for four of the more significant DES candidates are reported in \tabref{properties}.
Like the previously known ultra-faint dwarfs, the new DES systems are old and metal-poor \citep[e.g.,][]{2014ApJ...796...91B}.
The latter fitting procedure has also been applied to non-extinction corrected magnitudes to independently validate the extinction values from \citet{1998ApJ...500..525S}. 

\begin{\tabletype}{l ccc ccc}
\tablecolumns{13}
\tablewidth{0pt}
\tabletypesize{\tiny}
\tablecaption{Properties of DES satellite galaxy candidates\label{tab:properties}}
\tablehead{
Name & Distance & $M_{*}$ & $M_{V}$ & $r_{\rm 1/2}$ & $\log_{10}(\tau)$ & $Z$ \\
     &  (kpc)   & $(10^3 M_{\odot})$ & (mag)  &  (pc)      & $\log_{10}(\Gyr)$ & }
\startdata
DES\,J0335.6-5403 (Ret\,II)  & 32  & $2.6^{+0.2}_{-0.2}$    & $-3.6\pm0.1$   & $55^{+5}_{-5}$    & $10.08\pm0.21$ & $<0.0003$ \\
DES\,J0344.3-4331 (Eri\,II)  & 330 & $83^{+17}_{-14}$ & $-7.4\pm0.1$   & $172^{+57}_{-57}$ & $10.10\pm0.23$ & $<0.0006$ \\
DES\,J2251.2-5836 (Tuc\,II)  & 58  & $3^{+7}_{-1}$    & $-3.9\pm0.2$   & $120^{+30}_{-30}$ & -- & -- \\
DES\,J0255.4-5406 (Hor\,I)   & 87  & $2.4^{+1.4}_{-0.7}$    & $-3.5\pm0.3$   & $60^{+76}_{-30}$  & $9.96\pm0.21$ & $<0.0005$ \\
DES\,J2108.8-5109 (Ind\,I)   & 69  & $0.8^{+0.4}_{-0.4}$ & $-2.2\pm0.5$   & $12^{+2}_{-2}$ & -- & -- \\
DES\,J0443.8-5017 (Pic\,I)   & 126 & $2.8^{+5.0}_{-1.7}$    & $-3.7\pm0.4$   & $43^{+153}_{-21}$ & $10.00\pm0.16$ & $<0.0004$ \\
DES\,J2339.9-5424 (Phe\,II)  & 95  & $2.8^{+1.2}_{-0.7}$    & $-3.7\pm0.4$   & $33^{+20}_{-11}$  & -- & -- \\
DES\,J0222.7-5217 (Eri\,III) & 95  & $0.9^{+0.9}_{-0.7}$ & $-2.4\pm0.6$   & $11^{+8}_{-5}$ & -- & -- \\
\tableline

\enddata
\tablecomments{Uncertainties are calculated from the the highest density interval containing 90\% of the posterior distribution. Stellar masses are computed for a Chabrier initial mass function.}
\end{\tabletype}

\subsection{Review of Individual Candidates}
\label{sec:individual}

\begin{figure*}
  \includegraphics[width=\textwidth]{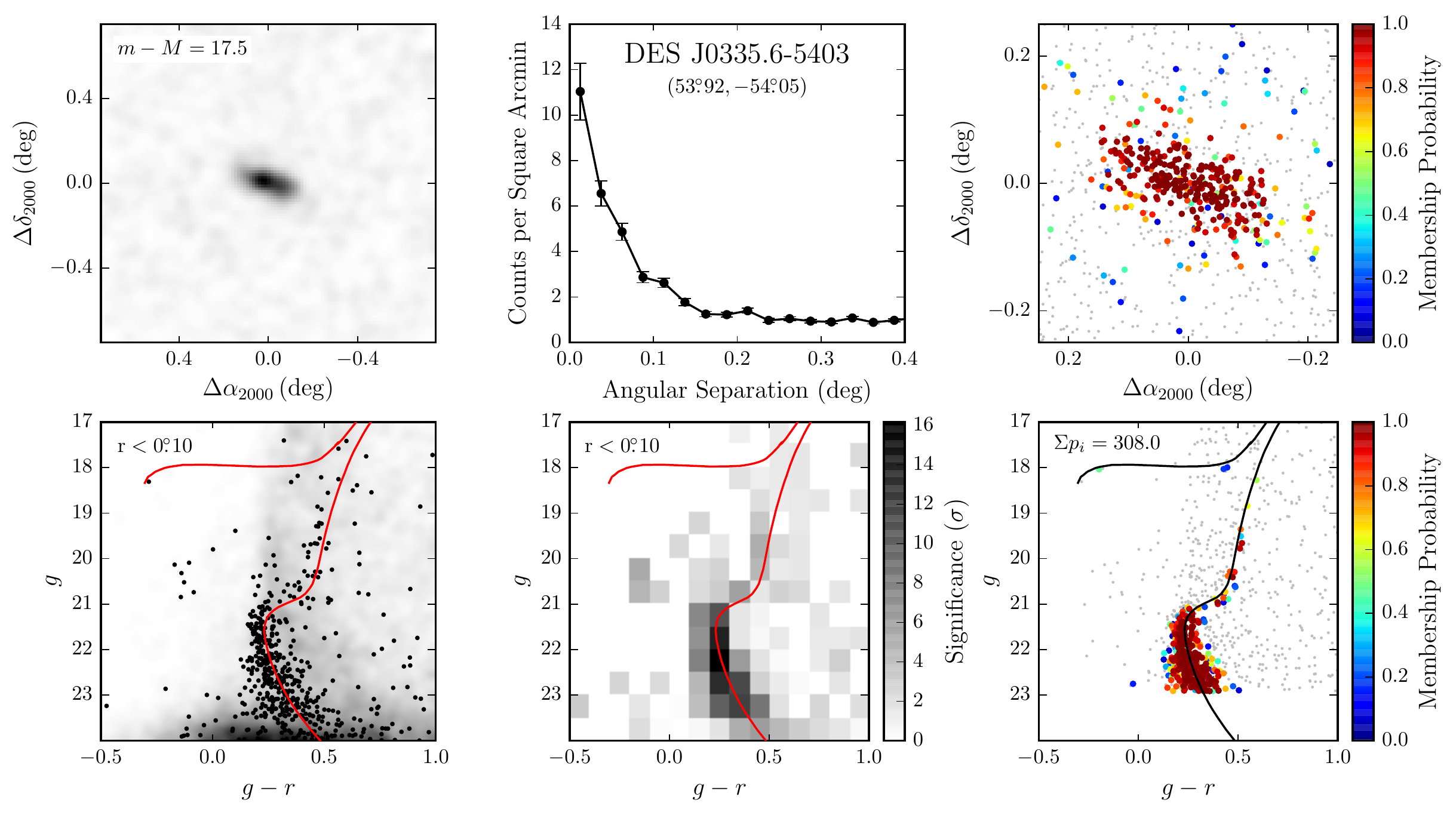}
  \caption{Stellar density and color-magnitude diagrams for \retII. 
  \textit{Top left}: Spatial distribution of stars with $g < 24 \magn$ that are within $0.1 \magn$ of the isochrone displayed in the lower panels. The field of view is $1\fdg5 \times 1\fdg5$ centered on the candidate and the stellar distribution has been smoothed with a Gaussian kernel with standard deviation $0\fdg027$.
  \textit{Top center}: Radial distribution of stars with $g - r < 1 \magn$ and $g < 24 \magn$.
   \textit{Top right}: Spatial distribution of stars with high membership probabilities within a $0\fdg5 \times 0\fdg5$ field of view. Small gray points indicate stars with membership probability less than 5\%.
   \textit{Bottom left}: The color-magnitude distribution of stars within $0.1\deg$ of the centroid are indicated with individual points. The density of the field within a $1\deg$ annulus is represented by the background two-dimensional histogram in grayscale. The red curve shows a representative isochrone for a stellar population with $\age = 13.5\Gyr$ and $\metal = 0.0001$ located at the best-fit distance modulus listed in the upper left panel. 
   \textit{Bottom center}: Binned significance diagram representing the Poisson probability of detecting the observed number of stars within the central 0.1 deg for each bin of the color-magnitude space given the local field density.
   \textit{Bottom right}: Color-magnitude distribution of high membership probability stars.
  }
\label{fig:cmd_retII}
\end{figure*}

Brief comments on the individual galaxy candidates are provided below, and spatial maps and color-magnitude diagrams for each candidate are provided in Figures~\ref{fig:cmd_retII} -- \ref{fig:cmd_eriIII}.
The rightmost panels of these figures show the satellite membership probabilities of individual stars that are assigned by the likelihood fit using a single representative isochrone with $\age = 13.5$~Gyr and $\metal = 0.0001$.
Stars with high membership probabilities contribute most to the statistical significance of each candidate.
The constellation designation, should these candidates be confirmed as dwarf galaxies, is listed in parenthesis.

\begin{itemize}

\item {\bf \retII} (Reticulum~II, \figref{cmd_retII}): As the nearest and most significant candidate, \retII is highly conspicuous in the Y1A1 stellar density maps, with $\roughly 300$ member stars brighter than $g \sim 23 \magn$.  
In fact, an over-density of faint stars at this position is even visible in the much shallower Digitized Sky Survey images, although it was not detected by \citet{2007AJ....133..715W} and other searches of photographic material.  
Note that like the previously known ultra-faint dwarfs, \retII very likely contains several blue horizontal branch stars identified by the likelihood procedure, two of which are relatively far from the center of the object.  
Given its luminosity, radius, and ellipticity, \retII is almost certainly a dwarf galaxy rather than a globular cluster.  
As illustrated in \figref{size_luminosity}, it is significantly more extended than any known faint globular cluster, and its elongated shape would also make it an extreme outlier from the Milky Way cluster population.  
Among known dwarfs, \retII appears quite comparable to Ursa Major II \citep{2006ApJ...650L..41Z,2010AJ....140..138M}.  
\retII is only $\roughly23\kpc$ from the LMC, and measurements of its radial velocity and proper motion will provide strong clues as to whether it originated as a Milky Way satellite or fell into the Milky Way halo as part of a Magellanic group.

\begin{figure*}
  \includegraphics[width=1.\textwidth]{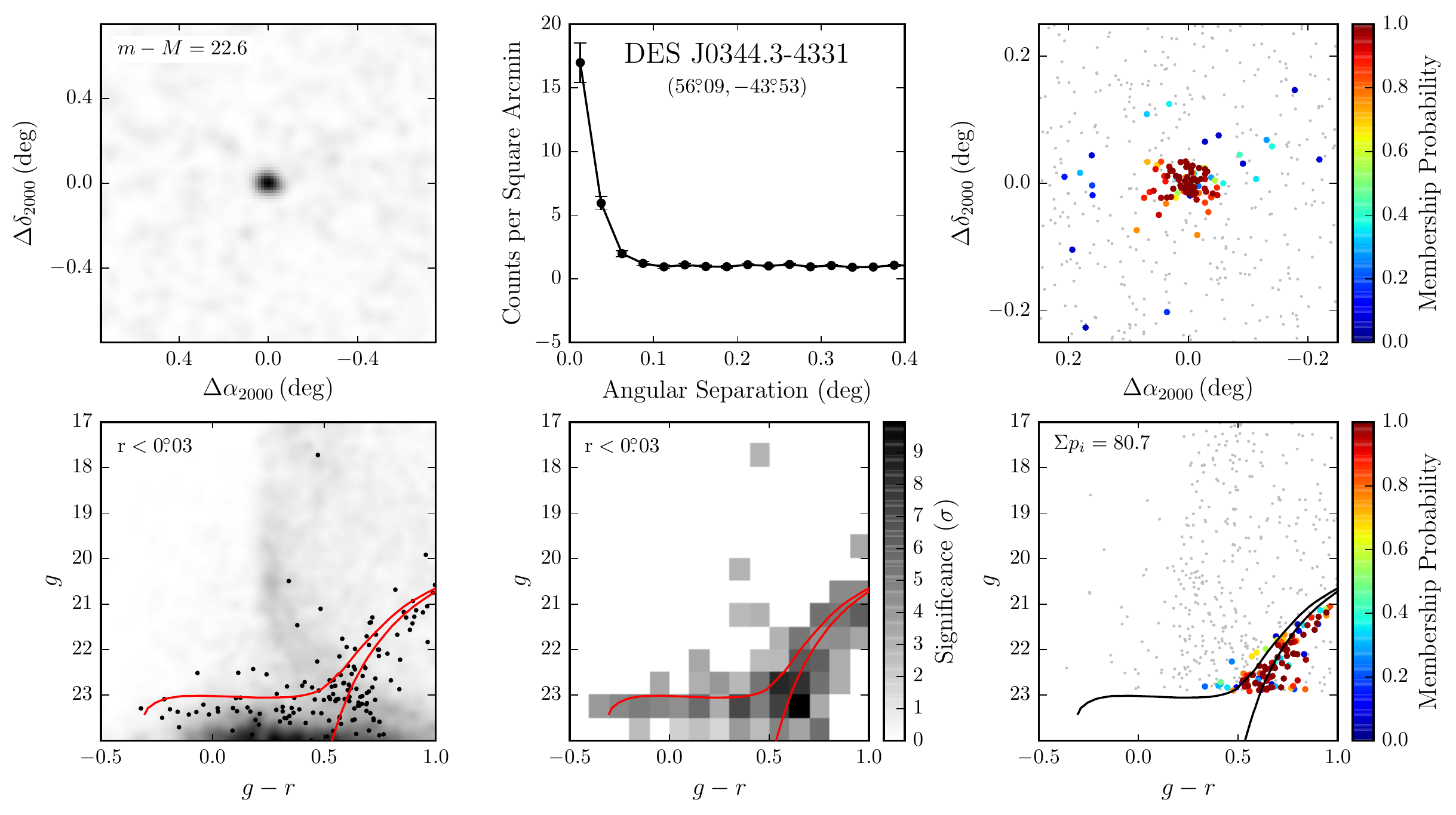}
  \caption{Analogous to \figref{cmd_retII} but for \eriII. A large number of stars, including several probable horizontal branch members, are present at magnitudes fainter than the $g < 23 \magn$ threshold of our likelihood analysis. This threshold was set by the rapidly decreasing stellar completeness at fainter magnitudes. However, it is likely that extending to fainter magnitudes would cause the best-fit distance modulus of \eriII to increase. Better constraints on the properties of \eriII require the stellar completeness to be robustly quantified in this regime.
  }
\label{fig:cmd_eriII}
\end{figure*}

\item {\bf \eriII} (Eridanus~II, \figref{cmd_eriII}): At a distance of ${>}\,330\kpc$, \eriII is nearly a factor of three more distant than any known outer halo globular cluster, and its half-light radius of $\roughly 170 \pc$ is inconsistent with the sizes of globular clusters.  
It is therefore very likely that this object is a new dwarf galaxy.
The color-magnitude diagram of \eriII closely resembles that of another distant Milky Way satellite, Canes Venatici~I, with a well-populated horizontal branch covering a wide range of colors \citep{2006ApJ...643L.103Z,2008ApJ...672L..13M,2012ApJ...744...96O}.
Its large distance places \eriII in a very intriguing range of parameter space for studying the quenching and loss of gas in dwarf galaxies.  
As has been known for many years, dwarf galaxies within $\sim250\kpc$ of the Milky Way and M31 are almost exclusively early-type galaxies with no gas or recent star formation, while dwarfs beyond that limit often have irregular morphologies, contain gas, and/or are still forming stars \citep[\eg,][]{1974Natur.252..111E,2000ApJ...541..675B,2009ApJ...696..385G,2014ApJ...795L...5S}.  
The next most distant Milky Way dwarf galaxy, Leo~T, is gas-rich and has hosted recent star formation \citep{2008MNRAS.384..535R,2008ApJ...680.1112D}; deeper optical and \ion{H}{1} imaging to search for neutral gas and young stars in \eriII has the potential to provide new insight into how gas is stripped from low-mass dwarfs and reveal the minimum mass for maintaining star formation over many \Gyr.  
A radial velocity measurement will also shed light on if \eriII has already passed close to the Milky Way or whether it is infalling for the first time.  
Given its distance, it is unlikely to be associated with the Magellanic Clouds.  
Like \retII, \eriII is clearly detected in Digitized Sky Survey images dating as far back as 1976.

\begin{figure*}
  \includegraphics[width=1.\textwidth]{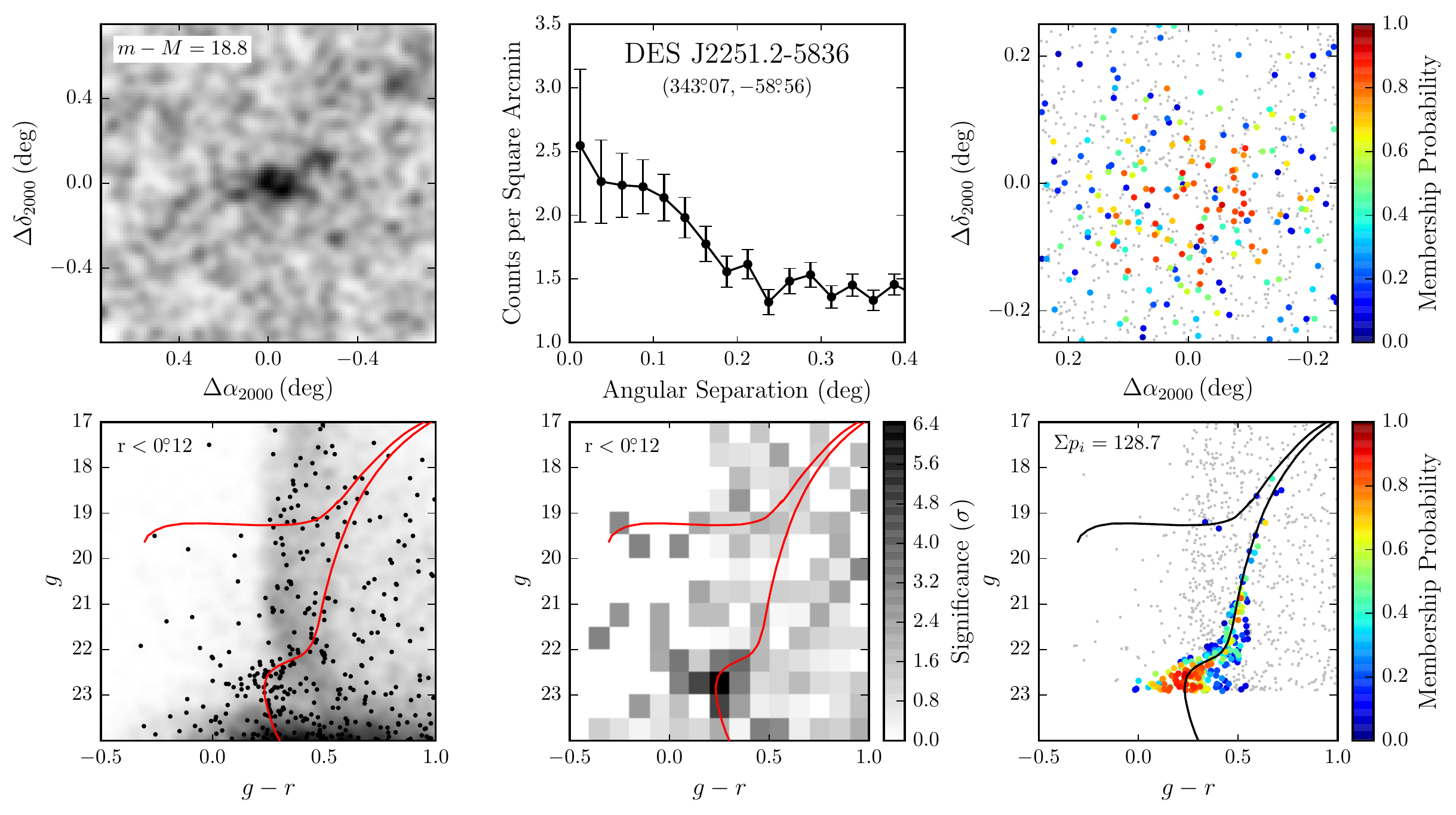}
  \caption{Analogous to \figref{cmd_retII} but for \tucII.
  }
\label{fig:cmd_tucII}
\end{figure*}

\item {\bf \tucII} (Tucana~II, \figref{cmd_tucII}): \tucII is the third new satellite with a large enough size ($120\pc$) to be tentatively identified as a dwarf galaxy with DES multi-band photometry alone. 
It has a similar luminosity to \retII, but is a much lower surface brightness system and is a factor of $\roughly 2$ farther away.  
\tucII is $\roughly 19\kpc$ from the LMC and $\roughly37\kpc$ from the SMC, making it a strong candidate for another member of the Magellanic group.
In the surface density map in the upper left panel of \figref{cmd_tucII}, the outer regions of \tucII appear elongated and distorted.  
However, these features are likely a result of noise rather than real distortions \citep{2008ApJ...684.1075M,2010AJ....140..138M}. 
The distribution of likely member stars in the upper right panel is much rounder.
The high detection significance of this object demonstrates the power of the likelihood analysis to simultaneously combine spatial and color-magnitude information.

\begin{figure*}
  \includegraphics[width=1.\textwidth]{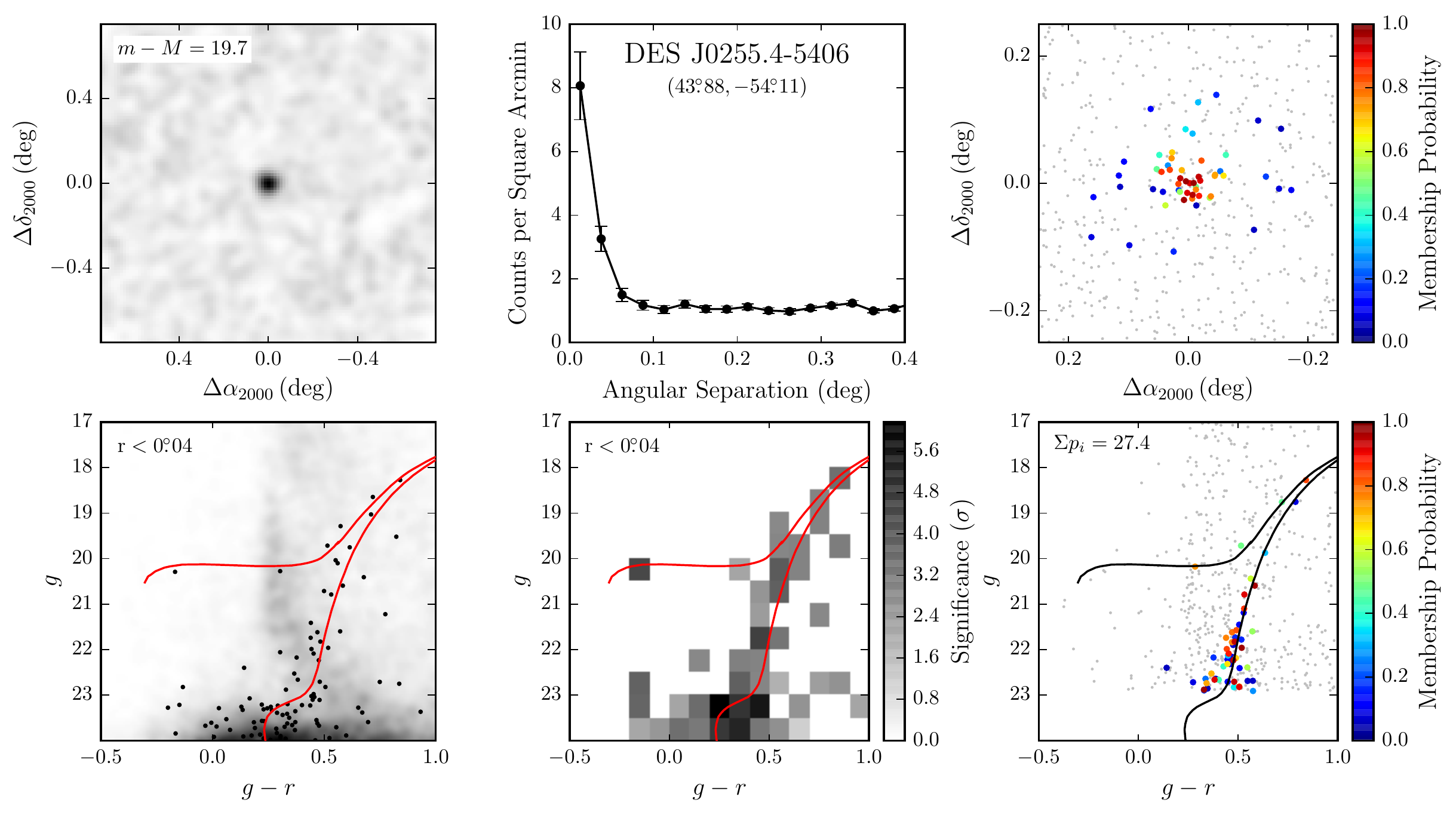}
  \caption{Analogous to \figref{cmd_retII} but for \horI.
  }
\label{fig:cmd_horI}
\end{figure*}

\item {\bf \horI} (Horologium~I, \figref{cmd_horI}): \horI, at a distance of $\roughly87\kpc$, has only a sparse population of RGB and HB stars visible in the DES photometry (a hint of the main sequence turnoff may be present at the detection limit).  
Given the small number of stars visible in the DES data, deeper imaging and spectroscopy will be needed to characterize the system more fully.  
Its Plummer radius of $\roughly60\pc$ establishes it as a likely dwarf galaxy, twice as extended as the largest globular clusters with comparable luminosities.  
\horI (perhaps along with \picI and \pheII; see below) is significantly farther from the Milky Way than any previously known dwarf galaxy with $M_{V} \lesssim -4$, suggesting that tidal stripping may not be needed to explain the low luminosities of the faintest dwarfs.
On its own, the Galactocentric distance is not necessarily a good indicator of the past importance of Galactic tides in shaping the photometric and spectroscopic properties of satellites. 
The most important factor is the peri-Galacticon distance, which is not yet known for the new satellites.
\horI is $\roughly 40\kpc$ away from the Magellanic Clouds and a factor $\sim2$ closer to them than to the Milky Way, making it a potential Magellanic satellite.
If it is (or was) associated with the Magellanic group, it is possible that tides from the LMC could have been relevant to its evolution.
A measurement of the systemic velocity will help clarify whether \horI is currently near apocenter, infalling, or associated with the Magellanic system.

\begin{figure*}
  \includegraphics[width=1.\textwidth]{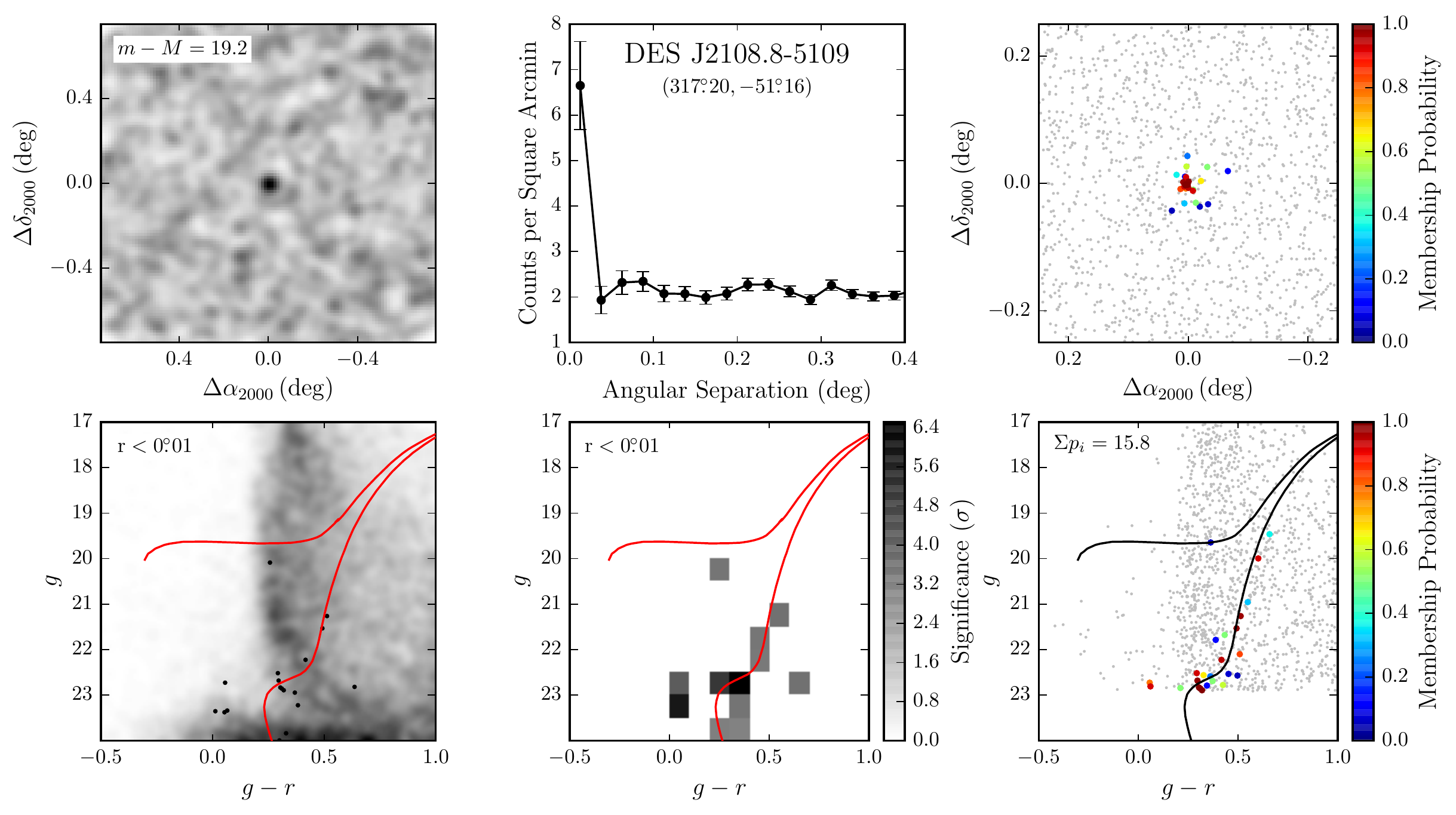}
  \caption{Analogous to \figref{cmd_retII} but for \indI.
  }
\label{fig:cmd_indI}
\end{figure*}

\item {\bf \indI} (Indus~I, \figref{cmd_indI}): We identify \indI with Kim~2 \citep{2015ApJ...803...63K}.
\indI is one of the faintest ($M_V \sim -2.2 \magn$) and most compact ($r_{h} \sim 12$~pc) of the satellites discussed here.
The object is visible in the coadded DES images. 
If it is a globular cluster, as argued by \citet{2015ApJ...803...63K}, \indI is fainter and more extended than most of the other outer halo clusters, such as AM~1, Eridanus, Pal~3, Pal~4, and Pal~14. 
\indI is $\roughly37$~kpc from the SMC, $\roughly55$~kpc from the LMC, and $\roughly69$~kpc from the much more massive Milky Way, so it is more likely a satellite of the Milky Way than of the Magellanic Clouds.

\begin{figure*}
  \includegraphics[width=1.\textwidth]{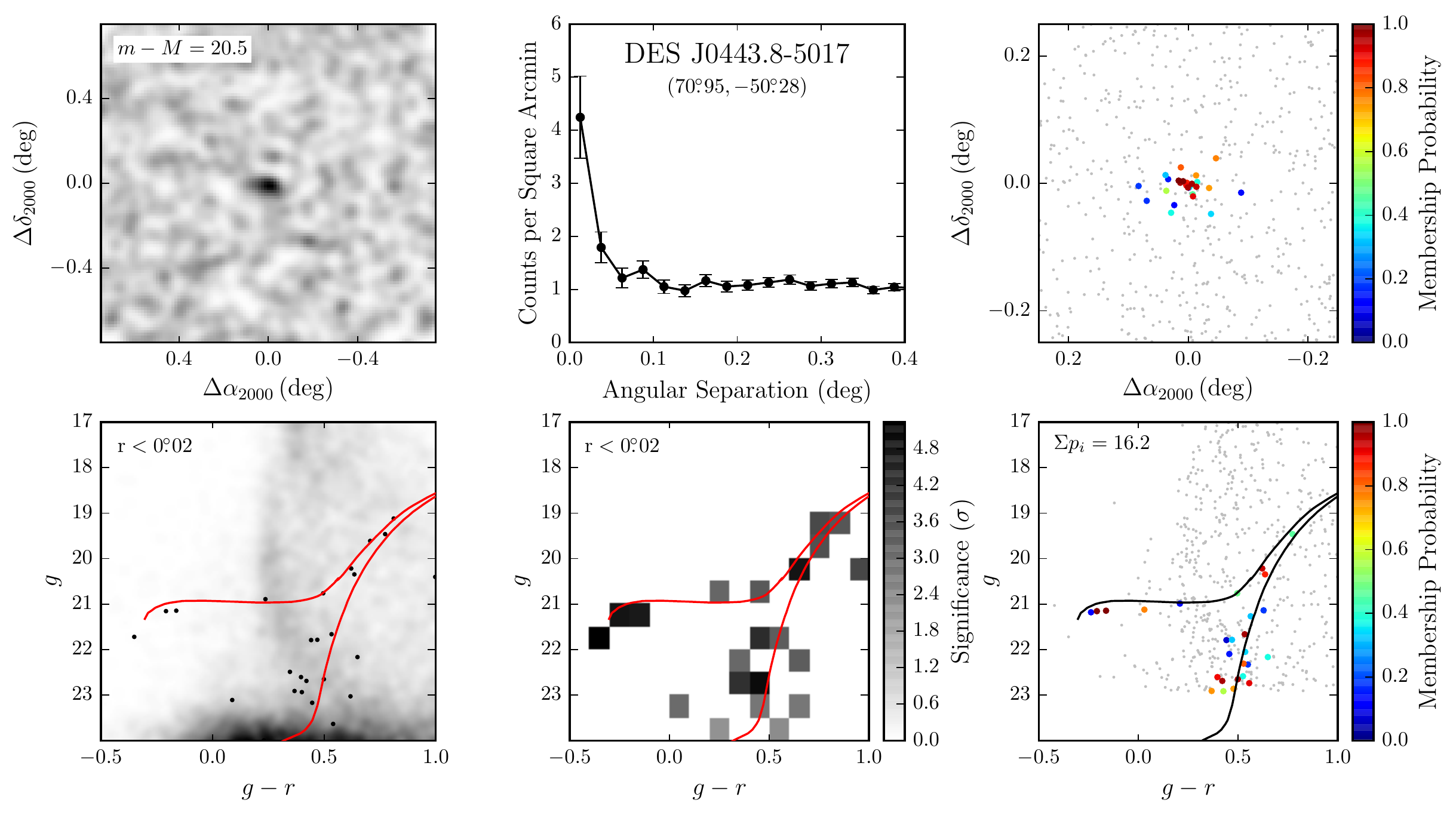}
  \caption{Analogous to \figref{cmd_retII} but for \picI.
  }
\label{fig:cmd_picI}
\end{figure*}

\item {\bf \picI} (Pictoris~I, \figref{cmd_picI}): \picI has a large enough radius to be a likely dwarf galaxy, but the uncertainty on the radius measurement is large enough to make it also consistent with the globular cluster population. 
\picI has a prominent blue horizontal branch and hints of an elliptical shape, but fewer member stars are detected in the DES data.
More accurate measurements of size and shape from deeper imaging, and kinematics and chemical abundances from spectroscopy, will be required to determine the nature of this object.
The large distance of \picI places it far enough behind the Magellanic Clouds that it is less likely to be a Magellanic satellite than many of the other new discoveries.

\begin{figure*}
  \includegraphics[width=1.\textwidth]{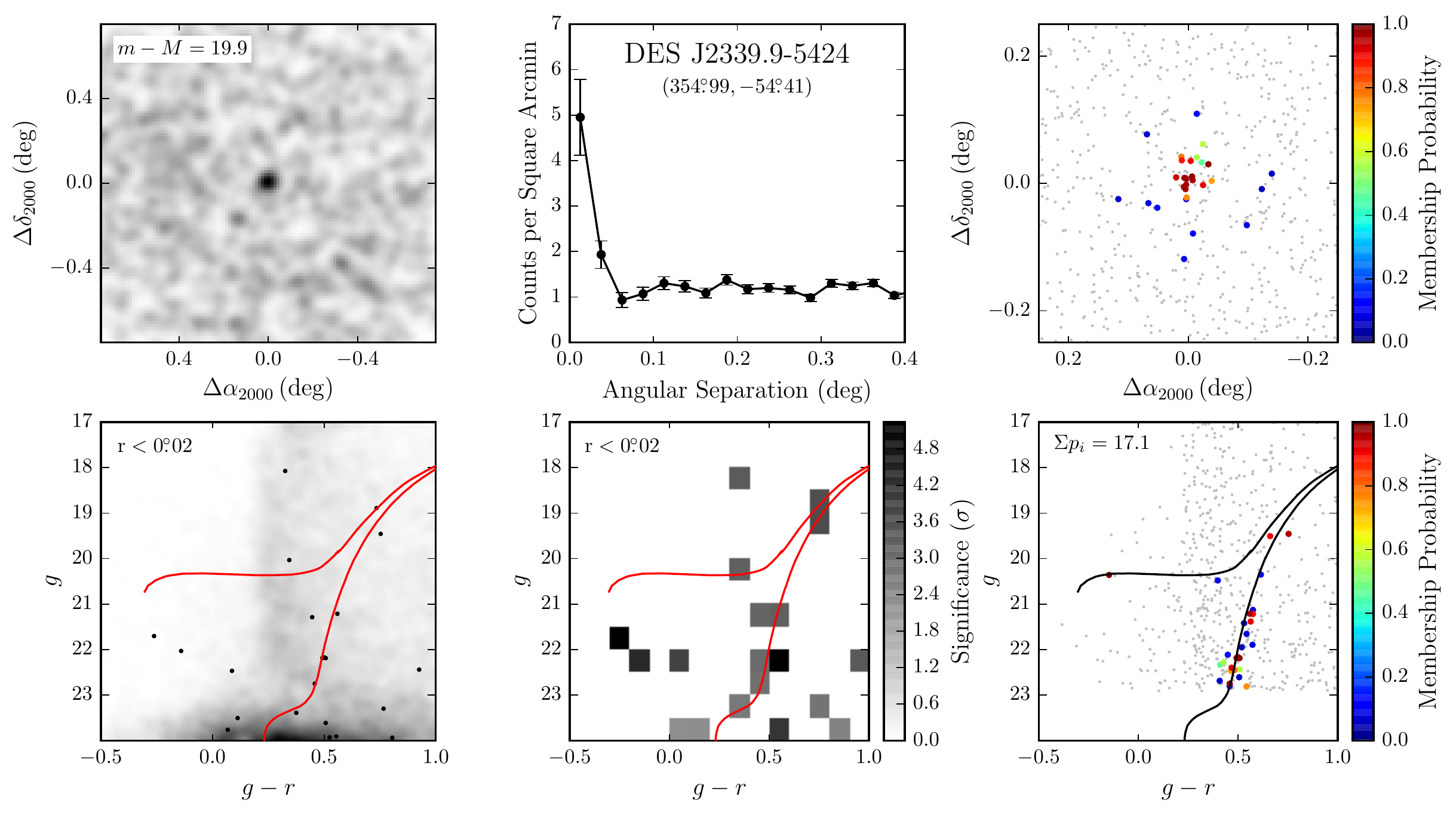}
  \caption{Analogous to \figref{cmd_retII} but for \pheII.
  }
\label{fig:cmd_pheII}
\end{figure*}

\item {\bf \pheII} (Phoenix~II, \figref{cmd_pheII}): \pheII is quite similar to \picI, but slightly smaller and closer.
Again, we cannot draw firm conclusions on its nature without additional data,   
At $\roughly 43\kpc$ from the SMC and $\roughly 65 \kpc$ from the LMC, it is unclear whether \pheII could plausibly be a Magellanic satellite.

\begin{figure*}
  \includegraphics[width=1.\textwidth]{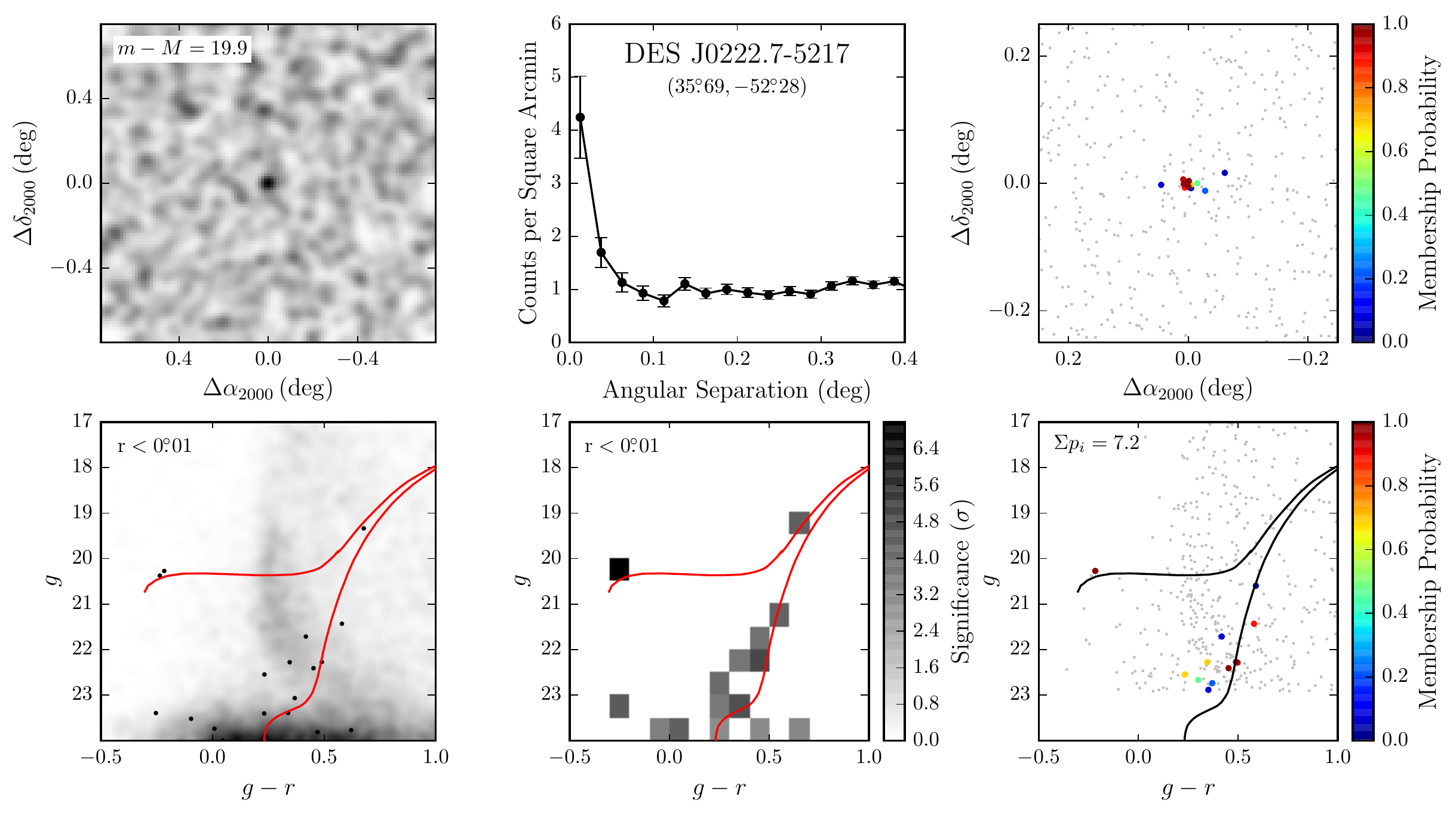}
  \caption{Analogous to \figref{cmd_retII} but for \eriIII.
  }
\label{fig:cmd_eriIII}
\end{figure*}

\item {\bf \eriIII} (Eridanus~III, \figref{cmd_eriIII}): \eriIII is the most compact of the newly discovered objects in both angular and physical units. 
Along with \indI, it is the most likely of the new discoveries to be a distant globular cluster rather than a dwarf galaxy.
However, they could also be consistent with an extension of the dwarf galaxy locus to fainter magnitudes and smaller sizes.
Even though it is one of the lowest luminosity system identified in the DES data so far, its compactness gives it a relatively high surface brightness, and like \indI and \picI, it is clearly visible in coadded images. 
However, only a handful of likely member stars are resolved at the depth of the Y1A1 data, and significantly deeper imaging will be needed to better constrain its physical properties and stellar population.

\end{itemize}

\subsection{Detection Completeness}
\label{sec:completeness}

\begin{\tabletype}{l ccc cccc}
\tablecolumns{13}
\tablewidth{0pt}
\tabletypesize{\tiny}
\tablecaption{Expected detection efficiencies for Milky Way companions in DES Y1A1\label{tab:efficiency}}
\tablehead{
Name & $M_V$ & Distance  & $r_h$    & Efficiency       & Efficiency        & Efficiency       & Efficiency \\
     &       & (kpc)     & $(\deg)$ & $(r_{\rm ext} = 0\fdg029)$ & $(r_{\rm ext} = 0\fdg057)$  & $(r_{\rm ext} = 0\fdg1)$ & $(r_{\rm ext} = r_h)$ }
\startdata
DES-J0335.6-5403 (Ret II) & -3.6 & 32 & 0.100 & 1.00 & 1.00 & 1.00 & 1.00 \\
DES-J0344.3-4331 (Eri II) & -7.4 & 330 & 0.030 & 1.00 & 1.00 & 1.00 & 1.00 \\
DES-J2251.2-5836 (Tuc II) & -3.9 & 58 & 0.120 & 0.25 & 0.98 & 1.00 & 1.00 \\
DES-J0255.4-5406 (Hor I) & -3.5 & 87 & 0.040 & 0.62 & 0.78 & 0.55 & 0.78 \\
DES-J2108.8-5109 (Ind I) & -2.2 & 69 & 0.010 & 0.96 & 0.69 & 0.18 & 0.97 \\
DES-J0443.8-5017 (Pic I) & -3.7 & 126 & 0.020 & 0.92 & 0.74 & 0.30 & 0.89 \\
DES-J2339.9-5424 (Phe II) & -3.7 & 95 & 0.020 & 1.00 & 0.96 & 0.74 & 0.99 \\
DES-J0222.7-5217 (Eri III) & -2.4 & 95 & 0.007 & 0.24 & 0.06 & 0.00 & 0.28 \\
\tableline
Segue 1 & -1.5 & 23 & 0.073 & 0.72 & 0.99 & 0.99 & 0.99 \\
Ursa Major II & -4.2 & 32 & 0.267 & 0.06 & 0.97 & 1.00 & 1.00 \\
Bootes II & -2.7 & 42 & 0.070 & 0.94 & 1.00 & 1.00 & 1.00 \\
Segue 2 & -2.5 & 35 & 0.057 & 1.00 & 1.00 & 1.00 & 1.00 \\
Willman 1 & -2.7 & 38 & 0.038 & 1.00 & 1.00 & 1.00 & 1.00 \\
Coma Berenices & -4.1 & 44 & 0.100 & 1.00 & 1.00 & 1.00 & 1.00 \\
Bootes III & -5.8 & 47 & 1.666 & 0.00 & 0.00 & 0.00 & 0.96 \\
Bootes I & -6.3 & 66 & 0.210 & 1.00 & 1.00 & 1.00 & 1.00 \\
Sextans & -9.3 & 86 & 0.463 & 1.00 & 1.00 & 1.00 & 1.00 \\
Ursa Major I & -5.5 & 97 & 0.188 & 0.00 & 0.30 & 0.90 & 0.98 \\
Hercules & -6.6 & 132 & 0.143 & 0.86 & 1.00 & 1.00 & 1.00 \\
Leo IV & -5.8 & 154 & 0.077 & 1.00 & 1.00 & 1.00 & 1.00 \\
Canes Venatici II & -4.9 & 160 & 0.027 & 1.00 & 1.00 & 1.00 & 1.00 \\
Leo V & -5.2 & 178 & 0.043 & 1.00 & 1.00 & 1.00 & 1.00 \\
Pisces II & -5.0 & 182 & 0.018 & 1.00 & 1.00 & 1.00 & 1.00 \\
Canes Venatici I & -8.6 & 218 & 0.148 & 1.00 & 1.00 & 1.00 & 1.00 \\
\tableline
\enddata
\tablecomments{Detection efficiencies are calculated from many realizations of satellites with the properties (luminosity $M_V$, distance, Plummer profile angular half-light radius $r_h$) of a given ultra-faint galaxy/candidate as they would have been observed in DES Y1A1. 
The simulated satellites are uniformly distributed throughout the SPT region of the Y1A1 footprint, excluding regions of high stellar density near to the LMC, i.e., $\roughly 1{,}600\deg^2$.
The rightmost columns list the detection efficiencies for extraction regions of different radii, $r_{\rm ext}$.
Here, a detection constitutes $>5\sigma$ stellar excess with $g < 23$ within the extraction region given the local density of the stellar field, after selecting stars that are consistent with the isochrone of an old and metal-poor stellar population at the satellite distance (i.e., following the map-based detection algorithm described in \secref{visual}).
The extraction region radii are choosen to reflect size scales used in the map-based search ($r_{\rm ext} = \{0\fdg029, 0\fdg057\}$), likelihood scan ($r_{\rm ext} = 0\fdg1$; \secref{likelihood}), and matched to the size of the satellite ($r_{\rm ext} = r_h$). Data for previously known satellites are taken from references compiled by \citet{2012AJ....144....4M}.}
\end{\tabletype}

Given that no additional ultra-faint Milky Way satellite galaxies have been confirmed outside of the SDSS DR7 footprint, despite the large areas of sky subsequently observed by SDSS and Pan-STARRS, it is interesting that multiple candidates have been found within the comparatively small area explored by DES thus far.\footnote{The classification of PSO~J174.0675-10.8774 / Crater~I as a globular cluster or dwarf galaxy is currently ambiguous \citep{2014ApJ...786L...3L,2014MNRAS.441.2124B}.}
Without definite classifications, it is difficult to incorporate the DES candidates into constraints on the luminosity function of Milky Way satellite galaxies. 
However, it is still possible to quantify the sensitivity of the first-year DES search using simple semi-analytic estimates of the completeness.

First, we calculated the probability that each new satellite could have been detected in the Y1A1 data.
We began by generating a large number of realizations of each galaxy candidate distributed uniformly over the Y1A1 footprint.
The candidates were modeled using radially symmetric Plummer profiles and the realizations included shot noise due to the limited number of stars expected to be in the observable magnitude range of DES.
We then applied the simple map-based detection algorithm described in \secref{visual} to evaluate the detection efficiency.
To be ``detected'', the satellite must possess at least 10 stars brighter than our imposed magnitude limit ($g < 23$) and a large enough surface brightness to pass the visual search selection criteria.
Specifically, we considered extraction of varying sizes and computed the Poisson probability of detecting $n_{\rm satellite} + n_{\rm field}$ stars when expecting $n_{\rm field}$ stars based on the local field surface density.
We tested extraction regions with sizes corresponding to the pixel areas in the map-based search algorithm ($r_{\rm ext} = \{ 0\fdg029, 0\fdg057\}$; \secref{visual}) and the kernel size from the likelihood scan ($r_{\rm ext} = 0\fdg1$; \secref{likelihood}), as well as an extraction radius set to the angular half-light radius of the simulated satellite.
When computing the local field density, we selected only stars along the isochrone at the distance of the satellite with $\age = 12\Gyr$ and $\metal = 0.0002$ (see \secref{visual}).
\tabref{efficiency} summarizes the expected detection efficiencies for the DES candidates when applying a $5 \sigma$ statistical significance threshold, as in our seed selection procedure for the map-based search.
The results show that all of the DES candidates would have been identified over a substantial fraction of the Y1A1 footprint with non-negligible probability, and for several candidates, near certainty.

\tabref{efficiency} also shows that the detection efficiency is sensitive to the size of the extraction region.
Extended systems such as \tucII are unlikely to found using the smallest extraction regions considered here, whereas the reverse is true for compact systems such as \eriIII and \indI.
This size dependence accounts for the low significance of the two most compact candidates, \eriIII and \indI, in the likelihood scan (\tabref{detection}).
After allowing their spatial extensions to be fit, the detection significances of these candidates increase to a level well above the our imposed threshold.

For comparison, \tabref{efficiency} also provides detection efficiency estimates for previously known ultra-faint galaxies (assuming they were located in the SPT region of Y1A1 instead of their actual locations).
We find that all of the SDSS ultra-faint galaxies, with the exception of the highly extended Bo\"{o}tes~III, could have been readily detected in Y1A1.
We attribute these high detection efficiencies to the deeper imaging of DES relative to SDSS and note that \tucII, \horI, \indI, and \eriIII have a substantially reduced detection probability when the magnitude limit is raised to $r<22\magn$, comparable to the stellar completeness limit of SDSS.

\begin{figure*}
  \includegraphics[width=\textwidth]{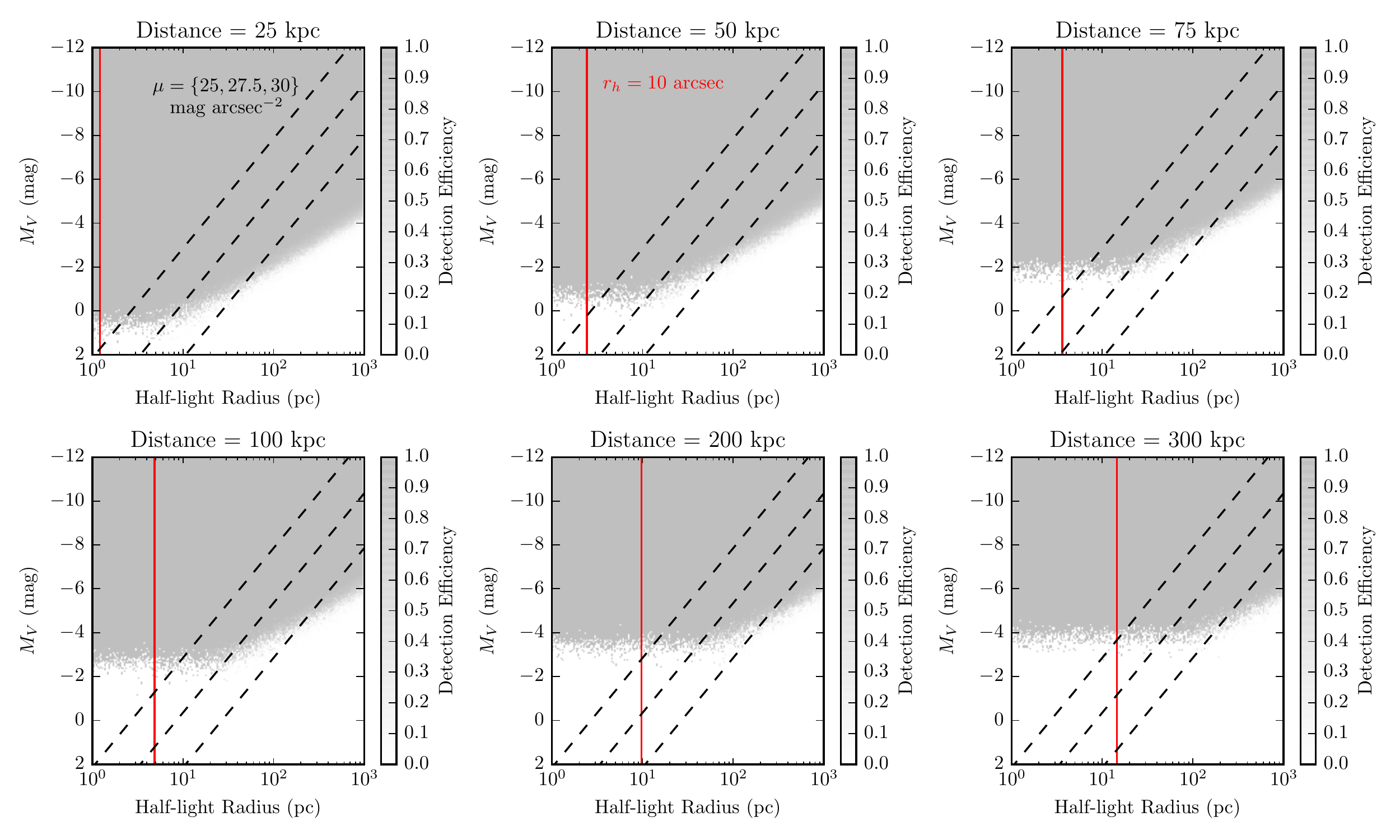}
  \caption{Sensitivity of the our Y1A1 search expressed as the detection efficiency as a function of satellite luminosity, physical size, and heliocentric distance. 
  Each panel corresponds to a different heliocentric distance.
  Contours of constant surface density are indicated with dashed black lines: $\mu$ = \{25, 27.5, 30\} mag arcsec$^{-2}$.
}
\label{fig:completeness}
\end{figure*}

Our Y1A1 search sensitivity can be quantified in a more general way by considering an ensemble of satellites spanning a range of luminosities, physical sizes, and heliocentric distances.
\figref{completeness} presents the discovery potential of our Y1A1 search expressed as the detection efficiency with respect to these galaxy properties, estimated by the same method described above with $r_{\rm ext} = r_{h}$.
Nearby, luminous, and compact objects have a high probability of being significantly detected whereas objects that are more distant, faint, and extended are less likely to be found.
The detection threshold in the plane of physical size and luminosity is nearly parallel to contours of constant surface brightness, and is weakly dependent on the distance, provided that a sufficient number of stars are detected.

Since the Y1A1 search procedure described in \secref{methods} is a combination of the map-based and likelihood-based search techniques, the actual completeness of our search is likely slightly higher than estimated here.
We expect the likelihood method to be more sensitive to extended low surface brightness systems because it combines spatial and color-magnitude information simultaneously.
As the depth of DES imaging increases (2 to 4 tilings in Y1A1 compared to 10 tilings planned after 5 years) and more advanced techniques are applied to separate stars and galaxies at faint magnitudes \citep[\eg,][]{2013arXiv1306.5236S,2012ApJ...760...15F}, we anticipate that lower surface brightness satellites will become accessible.
Our present study is optimized for the detection of relatively compact ($r_0 \lesssim 0\fdg2$) and radially symmetric stellar over-densities.
The search for extended low surface brightness features in the stellar distribution will be the focus of future work.

\subsection{Total Number and Spatial Distribution of Milky Way Satellite Galaxies}

The discovery of \nobjs new dwarf galaxy candidates in $\roughly 1{,}600 \deg^{2}$ of Y1A1 not overlapping with SDSS Stripe 82 is consistent with expectations from the literature \citep{2008ApJ...688..277T,2011AJ....141..185R,PhysRevD.91.063515,2014ApJ...795L..13H}.
By empirically modeling the incompleteness of SDSS, \cite{2008ApJ...688..277T} predicted that 19 to 37 satellite galaxies could be found over the full DES footprint.
More recent estimates based on high-resolution $N$-body simulations \citep{2014ApJ...795L..13H} and semi-analytic galaxy formation models that include baryonic physics \citep{PhysRevD.91.063515} predict $\roughly10$ new detectable satellite galaxies in DES.
Large uncertainties are associated with each of these estimates due to weak constraints on the luminosity function in the ultra-faint regime.
Additionally, as noted in \secref{individual}, some of the DES candidates may be globular clusters or may be associated with the Magellanic Clouds.
In the latter case, it becomes more challenging to directly compare our results to the predictions above, which assume an isotropic distribution of Milky Way satellite galaxies.

A number of studies, beginning with \citet{1976MNRAS.174..695L}, note that many Milky Way satellite galaxies appear to be distributed on the sky along a great circle, indicating a planar three-dimensional structure rather than an ellipsoidal or spherical distribution. 
This great circle has a polar orientation relative to the disk of the Milky Way.
The discovery of most of the SDSS ultra-faint dwarfs in the north Galactic cap region increased the apparent significance of this alignment.  
However, since the primary region surveyed by SDSS is located in the direction of this so-called vast polar structure \citep{2012MNRAS.423.1109P}, the true anisotropy of the Milky Way satellite population is not yet clear.  
The next generation of deep wide-field surveys should be able to address this issue with wider sky coverage.

In this context, it is interesting to consider the locations of the eight new satellites reported here, which may increase the known Milky Way dwarf galaxy population by $\roughly30\%$.  
The thickness of the vast polar structure defined by \citet{2012MNRAS.423.1109P} is 29~kpc, and we find that the DES satellites have a dispersion of $\roughly 28\kpc$ from this plane. 
This result is perhaps not surprising given their proximity to the Magellanic Clouds, which played a large role in defining the original Lynden-Bell plane.  
In fact, the entire Y1A1 search area (with the exception of Stripe 82) is located quite close to the previously known plane of satellites.  
Thus, any satellite galaxies identified in this data set are necessarily close to the plane, and a selection of eight random positions within this area would likely have a similar dispersion relative to the polar structure.  
A more quantitative characterization of the distribution of Milky Way satellites awaits the completion of the DES survey, including areas farther away from the vast polar structure, as well as future results from Pan-STARRS.

\section{Conclusions}
\label{sec:conclusions}

We report on the discovery of \nobjs new dwarf galaxy candidates associated with the Milky Way and/or Magellanic Clouds found in $\roughly 1{,}800 \deg^{2}$ of imaging data collected during the first year of DES.
These satellites span a wide range of absolute magnitudes ($-2.2$ to $-7.4 \magn$), physical sizes (10~pc to 170~pc), and heliocentric distances ($30\kpc$ to $330\kpc$).
The projected positions of the DES candidates are in close proximity to the Magellanic Clouds, and it is possible that some may be associated with the Magellanic system.

The nature of these systems cannot be conclusively determined with photometry alone.
However, judging from their low surface brightnesses, ellipticities, and/or large distances, it is likely that several are new dwarf galaxies, in particular, \retII, \eriII, and \tucII.
If spectroscopically confirmed, the DES candidates may become the first ultra-faint galaxies identified outside the SDSS footprint, and would significantly increase the population of Local Group galaxies in the southern hemisphere.
The proximity of \retII, at $\roughly 30 \kpc$, suggests that it may be an interesting target for indirect dark matter searches using gamma-ray telescopes \citep[\eg,][]{Ackermann:2013yva}.
The implications of these candidate galaxies for indirect dark matter searches are discussed in a separate paper \citep{Drlica-Wagner:2015xua}.

The second year of the DES survey was completed on 15 February 2015.
In addition to filling in regions of non-uniform coverage in the western portion of the Y1A1 footprint, the second season expands the DES survey to encompass over $4{,}000 \deg^2$.
The sensitivity to ultra-faint satellite galaxies achieved with first-year DES data already exceeds that of SDSS (\secref{completeness}), and over the next five years, DES is expected to make an important contribution to our understanding of the Milky Way environment.

During the preparation of this manuscript, we were sent an independent study by Koposov \etal using publicly released images from the first year of DES. 
\citet{Koposov:2015cua} identify nine candidate satellites of the Milky Way and/or Magellanic Clouds, including \nobjs that overlap with candidates presented here.
The candidate that Koposov \etal refer to as Grus I is located outside the Y1A1 footprint in a region that was observed during the first year of DES with good image quality, but which did not have sufficient coverage in all bands to enter the coaddition stage of the standard DESDM pipeline.
Therefore, the stars that comprise Grus~I are not in the coadd object catalog that was used for this analysis.

We note that we have not used the coordinates shared by Koposov \etal as seeds in our analysis, nor have we tuned our search algorithms based on knowledge of the candidates reported in their work.
The search methods presented here yield significance maps of the entire Y1A1 footprint, and the reported detections are the most significant points in the maps unassociated with known objects.
While our final choice of $5\sigma$ for the significance threshold for reportable galaxy candidates was made after our knowledge of the results from Koposov \etal, the threshold was chosen to provide as much timely information to the astronomical community as possible with minimal likelihood of false positives, rather than for agreement with the Koposov \etal detections.
We conclude that the independent discovery of these Milky Way companions by two separate teams using distinct object catalogs and search algorithms strengthens the case for follow-up by the astronomical community.

\section{Acknowledgments}
\label{sec:ackn}

We thank Sergey Koposov and collaborators for sending a copy of their submitted paper with their nine discoveries, and Helmut Jerjen for pointing out the association between Kim~2 and \indI.
Marla Geha provided useful comments on the presentation of these results.
KB and ADW thank Beth Willman for advice regarding the search for ultra-faint galaxies.
ADW thanks Ellen Bechtol for her generous hospitality during the preparation of this manuscript.
We acknowledge helpful suggestions from the anonymous referee.
This work made use of computational resources at the SLAC National Accelerator Laboratory and University of Chicago Research Computing Center.
This material is based upon work supported by the National Science Foundation under Grant Number (1138766).
ACR acknowledges financial support provided by the PAPDRJ CAPES/FAPERJ Fellowship.
AAP was supported by DOE grant DE-AC02-98CH10886 and by JPL, run by Caltech under a contract for NASA.

Funding for the DES Projects has been provided by the U.S. Department of Energy, the U.S. National Science Foundation, the Ministry of Science and Education of Spain, 
the Science and Technology Facilities Council of the United Kingdom, the Higher Education Funding Council for England, the National Center for Supercomputing 
Applications at the University of Illinois at Urbana-Champaign, the Kavli Institute of Cosmological Physics at the University of Chicago, 
the Center for Cosmology and Astro-Particle Physics at the Ohio State University,
the Mitchell Institute for Fundamental Physics and Astronomy at Texas A\&M University, Financiadora de Estudos e Projetos, 
Funda{\c c}{\~a}o Carlos Chagas Filho de Amparo {\`a} Pesquisa do Estado do Rio de Janeiro, Conselho Nacional de Desenvolvimento Cient{\'i}fico e Tecnol{\'o}gico and 
the Minist{\'e}rio da Ci{\^e}ncia, Tecnologia e Inova{\c c}{\~a}o, the Deutsche Forschungsgemeinschaft and the Collaborating Institutions in the Dark Energy Survey. 
The DES data management system is supported by the National Science Foundation under Grant Number AST-1138766.
The DES participants from Spanish institutions are partially supported by MINECO under grants AYA2012-39559, ESP2013-48274, FPA2013-47986, and Centro de Excelencia Severo Ochoa SEV-2012-0234, 
some of which include ERDF funds from the European Union.

The Collaborating Institutions are Argonne National Laboratory, the University of California at Santa Cruz, the University of Cambridge, Centro de Investigaciones En{\'e}rgeticas, 
Medioambientales y Tecnol{\'o}gicas-Madrid, the University of Chicago, University College London, the DES-Brazil Consortium, the University of Edinburgh, 
the Eidgen{\"o}ssische Technische Hochschule (ETH) Z{\"u}rich, 
Fermi National Accelerator Laboratory, the University of Illinois at Urbana-Champaign, the Institut de Ci{\`e}ncies de l'Espai (IEEC/CSIC), 
the Institut de F{\'i}sica d'Altes Energies, Lawrence Berkeley National Laboratory, the Ludwig-Maximilians Universit{\"a}t M{\"u}nchen and the associated Excellence Cluster Universe, 
the University of Michigan, the National Optical Astronomy Observatory, the University of Nottingham, The Ohio State University, the University of Pennsylvania, the University of Portsmouth, 
SLAC National Accelerator Laboratory, Stanford University, the University of Sussex, and Texas A\&M University.

\bibliographystyle{apj}
\bibliography{main}

\begin{thebibliography}{}
\expandafter\ifx\csname natexlab\endcsname\relax\def\natexlab#1{#1}\fi

\bibitem[{Abbott {et~al.}(2005)}]{Abbott:2005bi}
Abbott, T., {et~al.} 2005, arXiv:astro-ph/0510346

\bibitem[{Ackermann {et~al.}(2014)}]{Ackermann:2013yva}
Ackermann, M., {et~al.} 2014, \prd, 89, 042001

\bibitem[{{Ad{\'e}n} {et~al.}(2009){Ad{\'e}n}, {Feltzing}, {Koch}, {Wilkinson},
  {Grebel}, {Lundstr{\"o}m}, {Gilmore}, {Zucker}, {Belokurov}, {Evans}, \&
  {Faria}}]{2009A&A...506.1147A}
{Ad{\'e}n}, D., {Feltzing}, S., {Koch}, A., {et~al.} 2009, \aap, 506, 1147

\bibitem[{{Ahn} {et~al.}(2014){Ahn}, {Alexandroff}, {Allende Prieto}, {Anders},
  {Anderson}, {Anderton}, {Andrews}, {Aubourg}, {Bailey}, {Bastien}, \&
  et~al.}]{2014ApJS..211...17A}
{Ahn}, C.~P., {Alexandroff}, R., {Allende Prieto}, C., {et~al.} 2014, \apjs,
  211, 17

\bibitem[{{Balbinot} {et~al.}(2013){Balbinot}, {Santiago}, {da Costa}, {Maia},
  {Majewski}, {Nidever}, {Rocha-Pinto}, {Thomas}, {Wechsler}, \&
  {Yanny}}]{2013ApJ...767..101B}
{Balbinot}, E., {Santiago}, B.~X., {da Costa}, L., {et~al.} 2013, \apj, 767,
  101

\bibitem[{{Balbinot} {et~al.}(2015){Balbinot}, {Santiago}, {Girardi}, {Pieres},
  {da Costa}, {Maia}, {Gruendl}, {Walker}, {Yanny}, {Drlica-Wagner},
  {Benoit-Levy}, {Abbott}, {Allam}, {Annis}, {Bernstein}, {Bernstein},
  {Bertin}, {Brooks}, {Buckley-Geer}, {Rosell}, {Cunha}, {DePoy}, {Desai},
  {Diehl}, {Doel}, {Estrada}, {Evrard}, {Neto}, {Finley}, {Flaugher},
  {Frieman}, {Gruen}, {Honscheid}, {James}, {Kuehn}, {Kuropatkin}, {Lahav},
  {March}, {Marshall}, {Miller}, {Miquel}, {Ogando}, {Peoples}, {Plazas},
  {Scarpine}, {Schubnell}, {Sevilla-Noarbe}, {Smith}, {Soares-Santos},
  {Suchyta}, {Swanson}, {Tarle}, {Tucker}, {Wechsler}, \&
  {Zuntz}}]{2015MNRAS.449.1129B}
{Balbinot}, E., {Santiago}, B.~X., {Girardi}, L., {et~al.} 2015, \mnras, 449,
  1129

\bibitem[{{Belokurov} {et~al.}(2014){Belokurov}, {Irwin}, {Koposov}, {Evans},
  {Gonzalez-Solares}, {Metcalfe}, \& {Shanks}}]{2014MNRAS.441.2124B}
{Belokurov}, V., {Irwin}, M.~J., {Koposov}, S.~E., {et~al.} 2014, \mnras, 441,
  2124

\bibitem[{{Belokurov} {et~al.}(2006){Belokurov}, {Zucker}, {Evans},
  {Wilkinson}, {Irwin}, {Hodgkin}, {Bramich}, {Irwin}, {Gilmore}, {Willman},
  {Vidrih}, {Newberg}, {Wyse}, {Fellhauer}, {Hewett}, {Cole}, {Bell}, {Beers},
  {Rockosi}, {Yanny}, {Grebel}, {Schneider}, {Lupton}, {Barentine},
  {Brewington}, {Brinkmann}, {Harvanek}, {Kleinman}, {Krzesinski}, {Long},
  {Nitta}, {Smith}, \& {Snedden}}]{2006ApJ...647L.111B}
{Belokurov}, V., {Zucker}, D.~B., {Evans}, N.~W., {et~al.} 2006, \apjl, 647,
  L111

\bibitem[{{Belokurov} {et~al.}(2007){Belokurov}, {Zucker}, {Evans}, {Kleyna},
  {Koposov}, {Hodgkin}, {Irwin}, {Gilmore}, {Wilkinson}, {Fellhauer},
  {Bramich}, {Hewett}, {Vidrih}, {De Jong}, {Smith}, {Rix}, {Bell}, {Wyse},
  {Newberg}, {Mayeur}, {Yanny}, {Rockosi}, {Gnedin}, {Schneider}, {Beers},
  {Barentine}, {Brewington}, {Brinkmann}, {Harvanek}, {Kleinman}, {Krzesinski},
  {Long}, {Nitta}, \& {Snedden}}]{2007ApJ...654..897B}
---. 2007, \apj, 654, 897

\bibitem[{{Belokurov} {et~al.}(2008){Belokurov}, {Walker}, {Evans}, {Faria},
  {Gilmore}, {Irwin}, {Koposov}, {Mateo}, {Olszewski}, \&
  {Zucker}}]{2008ApJ...686L..83B}
{Belokurov}, V., {Walker}, M.~G., {Evans}, N.~W., {et~al.} 2008, \apjl, 686,
  L83

\bibitem[{{Belokurov} {et~al.}(2009){Belokurov}, {Walker}, {Evans}, {Gilmore},
  {Irwin}, {Mateo}, {Mayer}, {Olszewski}, {Bechtold}, \&
  {Pickering}}]{2009MNRAS.397.1748B}
---. 2009, \mnras, 397, 1748

\bibitem[{{Belokurov} {et~al.}(2010){Belokurov}, {Walker}, {Evans}, {Gilmore},
  {Irwin}, {Just}, {Koposov}, {Mateo}, {Olszewski}, {Watkins}, \&
  {Wyrzykowski}}]{2010ApJ...712L.103B}
---. 2010, \apjl, 712, L103

\bibitem[{{Bertin}(2011)}]{2011ASPC..442..435B}
{Bertin}, E. 2011, in Astronomical Society of the Pacific Conference Series,
  Vol. 442, Astronomical Data Analysis Software and Systems XX, ed. I.~N.
  {Evans}, A.~{Accomazzi}, D.~J. {Mink}, \& A.~H. {Rots}, 435

\bibitem[{{Bertin} \& {Arnouts}(1996)}]{1996A&AS..117..393B}
{Bertin}, E., \& {Arnouts}, S. 1996, \aaps, 117, 393

\bibitem[{{Blitz} \& {Robishaw}(2000)}]{2000ApJ...541..675B}
{Blitz}, L., \& {Robishaw}, T. 2000, \apj, 541, 675

\bibitem[{Box \& Tiao(1973)}]{box1973bayesian}
Box, G., \& Tiao, G. 1973, Bayesian inference in statistical analysis,
  Addison-Wesley series in behavioral science: quantitative methods
  (Addison-Wesley Pub. Co.)

\bibitem[{{Bressan} {et~al.}(2012){Bressan}, {Marigo}, {Girardi}, {Salasnich},
  {Dal Cero}, {Rubele}, \& {Nanni}}]{2012MNRAS.427..127B}
{Bressan}, A., {Marigo}, P., {Girardi}, L., {et~al.} 2012, \mnras, 427, 127

\bibitem[{{Brown} {et~al.}(2014){Brown}, {Tumlinson}, {Geha}, {Simon},
  {Vargas}, {VandenBerg}, {Kirby}, {Kalirai}, {Avila}, {Gennaro}, {Ferguson},
  {Mu{\~n}oz}, {Guhathakurta}, \& {Renzini}}]{2014ApJ...796...91B}
{Brown}, T.~M., {Tumlinson}, J., {Geha}, M., {et~al.} 2014, \apj, 796, 91

\bibitem[{{Carlin} {et~al.}(2009){Carlin}, {Grillmair}, {Mu{\~n}oz}, {Nidever},
  \& {Majewski}}]{2009ApJ...702L...9C}
{Carlin}, J.~L., {Grillmair}, C.~J., {Mu{\~n}oz}, R.~R., {Nidever}, D.~L., \&
  {Majewski}, S.~R. 2009, \apjl, 702, L9

\bibitem[{Carlstrom {et~al.}(2011)Carlstrom, Ade, Aird, Benson, Bleem,
  {et~al.}}]{Carlstrom:2009um}
Carlstrom, J., Ade, P., Aird, K., {et~al.} 2011, Publ.Astron.Soc.Pac., 123, 568

\bibitem[{{Chabrier}(2001)}]{2001ApJ...554.1274C}
{Chabrier}, G. 2001, \apj, 554, 1274

\bibitem[{{Chernoff}(1954)}]{chernoff1954likelihood}
{Chernoff}, H. 1954, Ann. Math. Stat., 25, 573

\bibitem[{Corwin(2004)}]{2004yCat.7239....0H}
Corwin, H.~G. 2004, VizieR Online Data Catalog, 7239, 0

\bibitem[{{de Jong} {et~al.}(2008){de Jong}, {Harris}, {Coleman}, {Martin},
  {Bell}, {Rix}, {Hill}, {Skillman}, {Sand}, {Olszewski}, {Zaritsky},
  {Thompson}, {Giallongo}, {Ragazzoni}, {DiPaola}, {Farinato}, {Testa}, \&
  {Bechtold}}]{2008ApJ...680.1112D}
{de Jong}, J.~T.~A., {Harris}, J., {Coleman}, M.~G., {et~al.} 2008, \apj, 680,
  1112

\bibitem[{{Desai} {et~al.}(2012){Desai}, {Armstrong}, {Mohr}, {Semler}, {Liu},
  {Bertin}, {Allam}, {Barkhouse}, {Bazin}, {Buckley-Geer}, {Cooper}, {Hansen},
  {High}, {Lin}, {Lin}, {Ngeow}, {Rest}, {Song}, {Tucker}, \&
  {Zenteno}}]{2012ApJ...757...83D}
{Desai}, S., {Armstrong}, R., {Mohr}, J.~J., {et~al.} 2012, \apj, 757, 83

\bibitem[{{Diehl} {et~al.}(2014)}]{2014SPIE.9149E..0VD}
{Diehl}, H.~T., {et~al.} 2014, Proc. SPIE, 9149, 0

\bibitem[{{Diehl} {et~al.}(2012)}]{DECamPaper2015}
{Diehl}, T., {et~al.} 2012, Physics Procedia, 37, 1332

\bibitem[{{Dolphin}(2002)}]{2002MNRAS.332...91D}
{Dolphin}, A.~E. 2002, \mnras, 332, 91

\bibitem[{Drlica-Wagner {et~al.}(2015)Drlica-Wagner, Albert, Bechtol,
  {et~al.}}]{Drlica-Wagner:2015xua}
Drlica-Wagner, A., Albert, A., Bechtol, K., {et~al.} 2015, submitted to ApJL,
  arXiv:1503.02632

\bibitem[{Edwards(1972)}]{edwards1972likelihood}
Edwards, A. 1972, Likelihood: An Account of the Statistical Concept of
  Likelihood and Its Application to Scientific Inference (Cambridge University
  Press)

\bibitem[{{Einasto} {et~al.}(1974){Einasto}, {Saar}, {Kaasik}, \&
  {Chernin}}]{1974Natur.252..111E}
{Einasto}, J., {Saar}, E., {Kaasik}, A., \& {Chernin}, A.~D. 1974, \nat, 252,
  111

\bibitem[{{Fadely} {et~al.}(2012){Fadely}, {Hogg}, \&
  {Willman}}]{2012ApJ...760...15F}
{Fadely}, R., {Hogg}, D.~W., \& {Willman}, B. 2012, \apj, 760, 15

\bibitem[{{Fadely} {et~al.}(2011){Fadely}, {Willman}, {Geha}, {Walsh},
  {Mu{\~n}oz}, {Jerjen}, {Vargas}, \& {Da Costa}}]{2011AJ....142...88F}
{Fadely}, R., {Willman}, B., {Geha}, M., {et~al.} 2011, \aj, 142, 88

\bibitem[{{Fisher}(1925)}]{1925PCPS...22..700F}
{Fisher}, R.~A. 1925, Proceedings of the Cambridge Philosophical Society, 22,
  700

\bibitem[{{Flaugher} {et~al.}(2015){Flaugher}, {Diehl}, {Honscheid}, {Abbott},
  {Alvarez}, {Angstadt}, {Annis}, {Antonik}, {Ballester}, {Beaufore},
  {Bernstein}, {Bernstein}, {Bigelow}, {Bonati}, {Boprie}, {Brooks},
  {Buckley-Geer}, {Campa}, {Cardiel-Sas}, {Castander}, {Castilla}, {Cease},
  {Cela-Ruiz}, {Chappa}, {Chi}, {Cooper}, {da Costa}, {Dede}, {Derylo},
  {DePoy}, {de Vicente}, {Doel}, {Drlica-Wagner}, {Eiting}, {Elliott}, {Emes},
  {Estrada}, {Fausti Neto}, {Finley}, {Flores}, {Frieman}, {Gerdes},
  {Gladders}, {Gregory}, {Gutierrez}, {Hao}, {Holland}, {Holm}, {Huffman},
  {Jackson}, {James}, {Jonas}, {Karcher}, {Karliner}, {Kent}, {Kessler},
  {Kozlovsky}, {Kron}, {Kubik}, {Kuehn}, {Kuhlmann}, {Kuk}, {Lahav}, {Lathrop},
  {Lee}, {Levi}, {Lewis}, {Li}, {Mandrichenko}, {Marshall}, {Martinez},
  {Merritt}, {Miquel}, {Munoz}, {Neilsen}, {Nichol}, {Nord}, {Ogando}, {Olsen},
  {Palio}, {Patton}, {Peoples}, {Plazas}, {Rauch}, {Reil}, {Rheault}, {Roe},
  {Rogers}, {Roodman}, {Sanchez}, {Scarpine}, {Schindler}, {Schmidt},
  {Schmitt}, {Schubnell}, {Schultz}, {Schurter}, {Scott}, {Serrano}, {Shaw},
  {Smith}, {Soares-Santos}, {Stefanik}, {Stuermer}, {Suchyta}, {Sypniewski},
  {Tarle}, {Thaler}, {Tighe}, {Tran}, {Tucker}, {Walker}, {Wang}, {Watson},
  {Weaverdyck}, {Wester}, {Woods}, \& {Yanny}}]{flaugher_2015_decam}
{Flaugher}, B., {Diehl}, H.~T., {Honscheid}, K., {et~al.} 2015, submitted to
  AJ, arXiv:1504.02900

\bibitem[{{Flaugher} {et~al.}(2010){Flaugher}, {Abbott}, {Annis}, {Antonik},
  {Bailey}, {Ballester}, {Bernstein}, {Bernstein}, {Bonati}, {Bremer},
  {Briones}, {Brooks}, {Buckley-Geer}, {Campa}, {Cardiel-Sas}, {Castander},
  {Castilla}, {Cease}, {Chappa}, {Chi}, {da Costa}, {DePoy}, {Derylo}, {De
  Vicente}, {Diehl}, {Doel}, {Estrada}, {Eiting}, {Elliott}, {Finley},
  {Frieman}, {Gaztanaga}, {Gerdes}, {Gladders}, {Guarino}, {Gutierrez},
  {Grudzinski}, {Hanlon}, {Hao}, {Holland}, {Honscheid}, {Huffman}, {Jackson},
  {Karliner}, {Kau}, {Kent}, {Krempetz}, {Krider}, {Kozlovsky}, {Kubik},
  {Kuehn}, {Kuhlmann}, {Kuk}, {Lahav}, {Lewis}, {Lin}, {Lorenzon}, {Marshall},
  {Mart{\'{\i}}nez}, {McKay}, {Merritt}, {Meyer}, {Miquel}, {Morgan}, {Moore},
  {Moore}, {Nord}, {Ogando}, {Olsen}, {Peoples}, {Plazas}, {Roe}, {Roodman},
  {Rossetto}, {Sanchez}, {Scarpine}, {Schalk}, {Schindler}, {Schmidt},
  {Schmitt}, {Schubnell}, {Schultz}, {Selen}, {Serrano}, {Shaw}, {Simaitis},
  {Slaughter}, {Smith}, {Spinka}, {Stefanik}, {Stuermer}, {Sypniewski},
  {Talaga}, {Tarle}, {Thaler}, {Tucker}, {Walker}, {Weaverdyck}, {Wester},
  {Woods}, {Worswick}, \& {Zhao}}]{2010SPIE.7735E..0DF}
{Flaugher}, B.~L., {Abbott}, T.~M.~C., {Annis}, J., {et~al.} 2010, Proc. SPIE,
  7735, 0

\bibitem[{{Foreman-Mackey} {et~al.}(2013){Foreman-Mackey}, {Hogg}, {Lang}, \&
  {Goodman}}]{2013PASP..125..306F}
{Foreman-Mackey}, D., {Hogg}, D.~W., {Lang}, D., \& {Goodman}, J. 2013, \pasp,
  125, 306

\bibitem[{{Geha} {et~al.}(2009){Geha}, {Willman}, {Simon}, {Strigari}, {Kirby},
  {Law}, \& {Strader}}]{2009ApJ...692.1464G}
{Geha}, M., {Willman}, B., {Simon}, J.~D., {et~al.} 2009, \apj, 692, 1464

\bibitem[{{G{\'o}rski} {et~al.}(2005){G{\'o}rski}, {Hivon}, {Banday},
  {Wandelt}, {Hansen}, {Reinecke}, \& {Bartelmann}}]{2005ApJ...622..759G}
{G{\'o}rski}, K.~M., {Hivon}, E., {Banday}, A.~J., {et~al.} 2005, \apj, 622,
  759

\bibitem[{{Grcevich} \& {Putman}(2009)}]{2009ApJ...696..385G}
{Grcevich}, J., \& {Putman}, M.~E. 2009, \apj, 696, 385

\bibitem[{{Grillmair}(2006)}]{2006ApJ...645L..37G}
{Grillmair}, C.~J. 2006, \apjl, 645, L37

\bibitem[{{Grillmair}(2009)}]{2009ApJ...693.1118G}
---. 2009, \apj, 693, 1118

\bibitem[{{Hargis} {et~al.}(2014){Hargis}, {Willman}, \&
  {Peter}}]{2014ApJ...795L..13H}
{Hargis}, J.~R., {Willman}, B., \& {Peter}, A.~H.~G. 2014, \apjl, 795, L13

\bibitem[{{Harris}(1996)}]{Harris96}
{Harris}, W.~E. 1996, \aj, 112, 1487

\bibitem[{He {et~al.}(2015)He, Bechtol, Hearin, \& Hooper}]{PhysRevD.91.063515}
He, C., Bechtol, K., Hearin, A.~P., \& Hooper, D. 2015, Phys. Rev. D, 91,
  063515

\bibitem[{{Irwin} {et~al.}(2007){Irwin}, {Belokurov}, {Evans}, {Ryan-Weber},
  {de Jong}, {Koposov}, {Zucker}, {Hodgkin}, {Gilmore}, {Prema}, {Hebb},
  {Begum}, {Fellhauer}, {Hewett}, {Kennicutt}, {Wilkinson}, {Bramich},
  {Vidrih}, {Rix}, {Beers}, {Barentine}, {Brewington}, {Harvanek},
  {Krzesinski}, {Long}, {Nitta}, \& {Snedden}}]{2007ApJ...656L..13I}
{Irwin}, M.~J., {Belokurov}, V., {Evans}, N.~W., {et~al.} 2007, \apjl, 656, L13

\bibitem[{{Jester} {et~al.}(2005){Jester}, {Schneider}, {Richards}, {Green},
  {Schmidt}, {Hall}, {Strauss}, {Vanden Berk}, {Stoughton}, {Gunn},
  {Brinkmann}, {Kent}, {Smith}, {Tucker}, \& {Yanny}}]{2005AJ....130..873J}
{Jester}, S., {Schneider}, D.~P., {Richards}, G.~T., {et~al.} 2005, \aj, 130,
  873

\bibitem[{{Kelly} {et~al.}(2014){Kelly}, {von der Linden}, {Applegate},
  {Allen}, {Allen}, {Burchat}, {Burke}, {Ebeling}, {Capak}, {Czoske},
  {Donovan}, {Mantz}, \& {Morris}}]{Kelly:2014}
{Kelly}, P.~L., {von der Linden}, A., {Applegate}, D.~E., {et~al.} 2014,
  \mnras, 439, 28

\bibitem[{{Kharchenko} {et~al.}(2013){Kharchenko}, {Piskunov}, {Schilbach},
  {R{\"o}ser}, \& {Scholz}}]{2013A&A...558A..53K}
{Kharchenko}, N.~V., {Piskunov}, A.~E., {Schilbach}, E., {R{\"o}ser}, S., \&
  {Scholz}, R.-D. 2013, \aap, 558, A53

\bibitem[{{Kim} \& {Jerjen}(2015)}]{2015ApJ...799...73K}
{Kim}, D., \& {Jerjen}, H. 2015, \apj, 799, 73

\bibitem[{{Kim} {et~al.}(2015){Kim}, {Jerjen}, {Milone}, {Mackey}, \& {Da
  Costa}}]{2015ApJ...803...63K}
{Kim}, D., {Jerjen}, H., {Milone}, A.~P., {Mackey}, D., \& {Da Costa}, G.~S.
  2015, \apj, 803, 63

\bibitem[{{Kirby} {et~al.}(2013){Kirby}, {Boylan-Kolchin}, {Cohen}, {Geha},
  {Bullock}, \& {Kaplinghat}}]{2013ApJ...770...16K}
{Kirby}, E.~N., {Boylan-Kolchin}, M., {Cohen}, J.~G., {et~al.} 2013, \apj, 770,
  16

\bibitem[{{Kleyna} {et~al.}(2005){Kleyna}, {Wilkinson}, {Evans}, \&
  {Gilmore}}]{2005ApJ...630L.141K}
{Kleyna}, J.~T., {Wilkinson}, M.~I., {Evans}, N.~W., \& {Gilmore}, G. 2005,
  \apjl, 630, L141

\bibitem[{{Klypin} {et~al.}(1999){Klypin}, {Kravtsov}, {Valenzuela}, \&
  {Prada}}]{1999ApJ...522...82K}
{Klypin}, A., {Kravtsov}, A.~V., {Valenzuela}, O., \& {Prada}, F. 1999, \apj,
  522, 82

\bibitem[{{Koch} {et~al.}(2009){Koch}, {Wilkinson}, {Kleyna}, {Irwin},
  {Zucker}, {Belokurov}, {Gilmore}, {Fellhauer}, \&
  {Evans}}]{2009ApJ...690..453K}
{Koch}, A., {Wilkinson}, M.~I., {Kleyna}, J.~T., {et~al.} 2009, \apj, 690, 453

\bibitem[{{Koposov} {et~al.}(2007){Koposov}, {de Jong}, {Belokurov}, {Rix},
  {Zucker}, {Evans}, {Gilmore}, {Irwin}, \& {Bell}}]{2007ApJ...669..337K}
{Koposov}, S., {de Jong}, J.~T.~A., {Belokurov}, V., {et~al.} 2007, \apj, 669,
  337

\bibitem[{{Koposov} {et~al.}(2008){Koposov}, {Belokurov}, {Evans}, {Hewett},
  {Irwin}, {Gilmore}, {Zucker}, {Rix}, {Fellhauer}, {Bell}, \&
  {Glushkova}}]{2008ApJ...686..279K}
{Koposov}, S., {Belokurov}, V., {Evans}, N.~W., {et~al.} 2008, \apj, 686, 279

\bibitem[{Koposov {et~al.}(2015)Koposov, Belokurov, Torrealba, \&
  Evans}]{Koposov:2015cua}
Koposov, S.~E., Belokurov, V., Torrealba, G., \& Evans, N.~W. 2015, accepted
  for publication in \apj, arXiv:1503.02079

\bibitem[{{Koposov} {et~al.}(2011){Koposov}, {Gilmore}, {Walker}, {Belokurov},
  {Wyn Evans}, {Fellhauer}, {Gieren}, {Geisler}, {Monaco}, {Norris}, {Okamoto},
  {Pe{\~n}arrubia}, {Wilkinson}, {Wyse}, \& {Zucker}}]{2011ApJ...736..146K}
{Koposov}, S.~E., {Gilmore}, G., {Walker}, M.~G., {et~al.} 2011, \apj, 736, 146

\bibitem[{{Laevens} {et~al.}(2014){Laevens}, {Martin}, {Sesar}, {Bernard},
  {Rix}, {Slater}, {Bell}, {Ferguson}, {Schlafly}, {Burgett}, {Chambers},
  {Denneau}, {Draper}, {Kaiser}, {Kudritzki}, {Magnier}, {Metcalfe}, {Morgan},
  {Price}, {Sweeney}, {Tonry}, {Wainscoat}, \& {Waters}}]{2014ApJ...786L...3L}
{Laevens}, B.~P.~M., {Martin}, N.~F., {Sesar}, B., {et~al.} 2014, \apjl, 786,
  L3

\bibitem[{{Lynden-Bell}(1976)}]{1976MNRAS.174..695L}
{Lynden-Bell}, D. 1976, \mnras, 174, 695

\bibitem[{{Martin} {et~al.}(2008{\natexlab{a}}){Martin}, {de Jong}, \&
  {Rix}}]{2008ApJ...684.1075M}
{Martin}, N.~F., {de Jong}, J.~T.~A., \& {Rix}, H.-W. 2008{\natexlab{a}}, \apj,
  684, 1075

\bibitem[{{Martin} {et~al.}(2007){Martin}, {Ibata}, {Chapman}, {Irwin}, \&
  {Lewis}}]{2007MNRAS.380..281M}
{Martin}, N.~F., {Ibata}, R.~A., {Chapman}, S.~C., {Irwin}, M., \& {Lewis},
  G.~F. 2007, \mnras, 380, 281

\bibitem[{{Martin} {et~al.}(2008{\natexlab{b}}){Martin}, {Coleman}, {De Jong},
  {Rix}, {Bell}, {Sand}, {Hill}, {Thompson}, {Burwitz}, {Giallongo},
  {Ragazzoni}, {Diolaiti}, {Gasparo}, {Grazian}, {Pedichini}, \&
  {Bechtold}}]{2008ApJ...672L..13M}
{Martin}, N.~F., {Coleman}, M.~G., {De Jong}, J.~T.~A., {et~al.}
  2008{\natexlab{b}}, \apjl, 672, L13

\bibitem[{{McConnachie}(2012{\natexlab{a}})}]{McConnachie12}
{McConnachie}, A.~W. 2012{\natexlab{a}}, \aj, 144, 4

\bibitem[{{McConnachie}(2012{\natexlab{b}})}]{2012AJ....144....4M}
---. 2012{\natexlab{b}}, \aj, 144, 4

\bibitem[{{Mohr} {et~al.}(2012){Mohr}, {Armstrong}, {Bertin}, {Daues}, {Desai},
  {Gower}, {Gruendl}, {Hanlon}, {Kuropatkin}, {Lin}, {Marriner}, {Petravic},
  {Sevilla}, {Swanson}, {Tomashek}, {Tucker}, \& {Yanny}}]{2012SPIE.8451E..0DM}
{Mohr}, J.~J., {Armstrong}, R., {Bertin}, E., {et~al.} 2012, Proc. SPIE, 8451,
  84510D

\bibitem[{{Moore} {et~al.}(1999){Moore}, {Ghigna}, {Governato}, {Lake},
  {Quinn}, {Stadel}, \& {Tozzi}}]{1999ApJ...524L..19M}
{Moore}, B., {Ghigna}, S., {Governato}, F., {et~al.} 1999, \apjl, 524, L19

\bibitem[{{Mu{\~n}oz} {et~al.}(2006){Mu{\~n}oz}, {Carlin}, {Frinchaboy},
  {Nidever}, {Majewski}, \& {Patterson}}]{2006ApJ...650L..51M}
{Mu{\~n}oz}, R.~R., {Carlin}, J.~L., {Frinchaboy}, P.~M., {et~al.} 2006, \apjl,
  650, L51

\bibitem[{{Mu{\~n}oz} {et~al.}(2012{\natexlab{a}}){Mu{\~n}oz}, {Geha},
  {C{\^o}t{\'e}}, {Vargas}, {Santana}, {Stetson}, {Simon}, \&
  {Djorgovski}}]{2012ApJ...753L..15M}
{Mu{\~n}oz}, R.~R., {Geha}, M., {C{\^o}t{\'e}}, P., {et~al.}
  2012{\natexlab{a}}, \apjl, 753, L15

\bibitem[{{Mu{\~n}oz} {et~al.}(2010){Mu{\~n}oz}, {Geha}, \&
  {Willman}}]{2010AJ....140..138M}
{Mu{\~n}oz}, R.~R., {Geha}, M., \& {Willman}, B. 2010, \aj, 140, 138

\bibitem[{{Mu{\~n}oz} {et~al.}(2012{\natexlab{b}}){Mu{\~n}oz}, {Padmanabhan},
  \& {Geha}}]{2012ApJ...745..127M}
{Mu{\~n}oz}, R.~R., {Padmanabhan}, N., \& {Geha}, M. 2012{\natexlab{b}}, \apj,
  745, 127

\bibitem[{{Nilson}(1973)}]{1973ugcg.book.....N}
{Nilson}, P. 1973, {Uppsala general catalogue of galaxies}

\bibitem[{{Okamoto} {et~al.}(2012){Okamoto}, {Arimoto}, {Yamada}, \&
  {Onodera}}]{2012ApJ...744...96O}
{Okamoto}, S., {Arimoto}, N., {Yamada}, Y., \& {Onodera}, M. 2012, \apj, 744,
  96

\bibitem[{{Ortolani} {et~al.}(2013){Ortolani}, {Bica}, \&
  {Barbuy}}]{2013MNRAS.433.1966O}
{Ortolani}, S., {Bica}, E., \& {Barbuy}, B. 2013, \mnras, 433, 1966

\bibitem[{{Paust} {et~al.}(2014){Paust}, {Wilson}, \& {van
  Belle}}]{2014AJ....148...19P}
{Paust}, N., {Wilson}, D., \& {van Belle}, G. 2014, \aj, 148, 19

\bibitem[{{Pawlowski} {et~al.}(2012){Pawlowski}, {Pflamm-Altenburg}, \&
  {Kroupa}}]{2012MNRAS.423.1109P}
{Pawlowski}, M.~S., {Pflamm-Altenburg}, J., \& {Kroupa}, P. 2012, \mnras, 423,
  1109

\bibitem[{{Plummer}(1911)}]{1911MNRAS..71..460P}
{Plummer}, H.~C. 1911, \mnras, 71, 460

\bibitem[{{Rossetto} {et~al.}(2011){Rossetto}, {Santiago}, {Girardi},
  {Camargo}, {Balbinot}, {da Costa}, {Yanny}, {Maia}, {Makler}, {Ogando},
  {Pellegrini}, {Ramos}, {de Simoni}, {Armstrong}, {Bertin}, {Desai},
  {Kuropatkin}, {Lin}, {Mohr}, \& {Tucker}}]{2011AJ....141..185R}
{Rossetto}, B.~M., {Santiago}, B.~X., {Girardi}, L., {et~al.} 2011, \aj, 141,
  185

\bibitem[{{Ryan-Weber} {et~al.}(2008){Ryan-Weber}, {Begum}, {Oosterloo}, {Pal},
  {Irwin}, {Belokurov}, {Evans}, \& {Zucker}}]{2008MNRAS.384..535R}
{Ryan-Weber}, E.~V., {Begum}, A., {Oosterloo}, T., {et~al.} 2008, \mnras, 384,
  535

\bibitem[{{Rykoff} {et~al.}(2014){Rykoff}, {Rozo}, {Busha}, {Cunha},
  {Finoguenov}, {Evrard}, {Hao}, {Koester}, {Leauthaud}, {Nord}, {Pierre},
  {Reddick}, {Sadibekova}, {Sheldon}, \& {Wechsler}}]{2014ApJ...785..104R}
{Rykoff}, E.~S., {Rozo}, E., {Busha}, M.~T., {et~al.} 2014, \apj, 785, 104

\bibitem[{{Sakamoto} \& {Hasegawa}(2006)}]{2006ApJ...653L..29S}
{Sakamoto}, T., \& {Hasegawa}, T. 2006, \apjl, 653, L29

\bibitem[{{Schlafly} {et~al.}(2012){Schlafly}, {Finkbeiner}, {Juri{\'c}},
  {Magnier}, {Burgett}, {Chambers}, {Grav}, {Hodapp}, {Kaiser}, {Kudritzki},
  {Martin}, {Morgan}, {Price}, {Rix}, {Stubbs}, {Tonry}, \&
  {Wainscoat}}]{2012ApJ...756..158S}
{Schlafly}, E.~F., {Finkbeiner}, D.~P., {Juri{\'c}}, M., {et~al.} 2012, \apj,
  756, 158

\bibitem[{{Schlegel} {et~al.}(1998){Schlegel}, {Finkbeiner}, \&
  {Davis}}]{1998ApJ...500..525S}
{Schlegel}, D.~J., {Finkbeiner}, D.~P., \& {Davis}, M. 1998, \apj, 500, 525

\bibitem[{{Sevilla} {et~al.}(2011){Sevilla}, {Armstrong}, {Bertin}, {Carlson},
  {Daues}, {Desai}, {Gower}, {Gruendl}, {Hanlon}, {Jarvis}, {Kessler},
  {Kuropatkin}, {Lin}, {Marriner}, {Mohr}, {Petravick}, {Sheldon}, {Swanson},
  {Tomashek}, {Tucker}, {Yang}, {Yanny}, \& {for the DES
  Collaboration}}]{Sevilla:2011}
{Sevilla}, I., {Armstrong}, R., {Bertin}, E., {et~al.} 2011, ArXiv e-prints,
  arXiv:1109.6741

\bibitem[{{Simon} \& {Geha}(2007)}]{2007ApJ...670..313S}
{Simon}, J.~D., \& {Geha}, M. 2007, \apj, 670, 313

\bibitem[{{Simon} {et~al.}(2011){Simon}, {Geha}, {Minor}, {Martinez}, {Kirby},
  {Bullock}, {Kaplinghat}, {Strigari}, {Willman}, {Choi}, {Tollerud}, \&
  {Wolf}}]{2011ApJ...733...46S}
{Simon}, J.~D., {Geha}, M., {Minor}, Q.~E., {et~al.} 2011, \apj, 733, 46

\bibitem[{{Soumagnac} {et~al.}(2013){Soumagnac}, {Abdalla}, {Lahav}, {Kirk},
  {Sevilla}, {Bertin}, {Rowe}, {Annis}, {Busha}, {Da Costa}, {Frieman},
  {Gaztanaga}, {Jarvis}, {Lin}, {Percival}, {Santiago}, {Sabiu}, {Wechsler},
  {Wolz}, \& {Yanny}}]{2013arXiv1306.5236S}
{Soumagnac}, M.~T., {Abdalla}, F.~B., {Lahav}, O., {et~al.} 2013, ArXiv
  e-prints, arXiv:1306.5236

\bibitem[{{Spekkens} {et~al.}(2014){Spekkens}, {Urbancic}, {Mason}, {Willman},
  \& {Aguirre}}]{2014ApJ...795L...5S}
{Spekkens}, K., {Urbancic}, N., {Mason}, B.~S., {Willman}, B., \& {Aguirre},
  J.~E. 2014, \apjl, 795, L5

\bibitem[{{Swanson} {et~al.}(2008){Swanson}, {Tegmark}, {Hamilton}, \&
  {Hill}}]{2008MNRAS.387.1391S}
{Swanson}, M.~E.~C., {Tegmark}, M., {Hamilton}, A.~J.~S., \& {Hill}, J.~C.
  2008, \mnras, 387, 1391

\bibitem[{{Tollerud} {et~al.}(2008){Tollerud}, {Bullock}, {Strigari}, \&
  {Willman}}]{2008ApJ...688..277T}
{Tollerud}, E.~J., {Bullock}, J.~S., {Strigari}, L.~E., \& {Willman}, B. 2008,
  \apj, 688, 277

\bibitem[{{van den Bergh}(2008)}]{2008AJ....135.1731V}
{van den Bergh}, S. 2008, \aj, 135, 1731

\bibitem[{{Walker} {et~al.}(2009){Walker}, {Belokurov}, {Evans}, {Irwin},
  {Mateo}, {Olszewski}, \& {Gilmore}}]{2009ApJ...694L.144W}
{Walker}, M.~G., {Belokurov}, V., {Evans}, N.~W., {et~al.} 2009, \apjl, 694,
  L144

\bibitem[{{Walsh} {et~al.}(2007){Walsh}, {Jerjen}, \&
  {Willman}}]{2007ApJ...662L..83W}
{Walsh}, S.~M., {Jerjen}, H., \& {Willman}, B. 2007, \apjl, 662, L83

\bibitem[{{Walsh} {et~al.}(2009){Walsh}, {Willman}, \&
  {Jerjen}}]{2009AJ....137..450W}
{Walsh}, S.~M., {Willman}, B., \& {Jerjen}, H. 2009, \aj, 137, 450

\bibitem[{{Whiting} {et~al.}(2007){Whiting}, {Hau}, {Irwin}, \&
  {Verdugo}}]{2007AJ....133..715W}
{Whiting}, A.~B., {Hau}, G.~K.~T., {Irwin}, M., \& {Verdugo}, M. 2007, \aj,
  133, 715

\bibitem[{{Willman} {et~al.}(2011){Willman}, {Geha}, {Strader}, {Strigari},
  {Simon}, {Kirby}, {Ho}, \& {Warres}}]{2011AJ....142..128W}
{Willman}, B., {Geha}, M., {Strader}, J., {et~al.} 2011, \aj, 142, 128

\bibitem[{{Willman} \& {Strader}(2012)}]{2012AJ....144...76W}
{Willman}, B., \& {Strader}, J. 2012, \aj, 144, 76

\bibitem[{{Willman} {et~al.}(2005{\natexlab{a}}){Willman}, {Blanton}, {West},
  {Dalcanton}, {Hogg}, {Schneider}, {Wherry}, {Yanny}, \&
  {Brinkmann}}]{2005AJ....129.2692W}
{Willman}, B., {Blanton}, M.~R., {West}, A.~A., {et~al.} 2005{\natexlab{a}},
  \aj, 129, 2692

\bibitem[{{Willman} {et~al.}(2005{\natexlab{b}}){Willman}, {Dalcanton},
  {Martinez-Delgado}, {West}, {Blanton}, {Hogg}, {Barentine}, {Brewington},
  {Harvanek}, {Kleinman}, {Krzesinski}, {Long}, {Neilsen}, {Nitta}, \&
  {Snedden}}]{2005ApJ...626L..85W}
{Willman}, B., {Dalcanton}, J.~J., {Martinez-Delgado}, D., {et~al.}
  2005{\natexlab{b}}, \apjl, 626, L85

\bibitem[{{Wolf} {et~al.}(2010){Wolf}, {Martinez}, {Bullock}, {Kaplinghat},
  {Geha}, {Mu{\~n}oz}, {Simon}, \& {Avedo}}]{2010MNRAS.406.1220W}
{Wolf}, J., {Martinez}, G.~D., {Bullock}, J.~S., {et~al.} 2010, \mnras, 406,
  1220

\bibitem[{{Zucker} {et~al.}(2006{\natexlab{a}}){Zucker}, {Belokurov}, {Evans},
  {Kleyna}, {Irwin}, {Wilkinson}, {Fellhauer}, {Bramich}, {Gilmore}, {Newberg},
  {Yanny}, {Smith}, {Hewett}, {Bell}, {Rix}, {Gnedin}, {Vidrih}, {Wyse},
  {Willman}, {Grebel}, {Schneider}, {Beers}, {Kniazev}, {Barentine},
  {Brewington}, {Brinkmann}, {Harvanek}, {Kleinman}, {Krzesinski}, {Long},
  {Nitta}, \& {Snedden}}]{2006ApJ...650L..41Z}
{Zucker}, D.~B., {Belokurov}, V., {Evans}, N.~W., {et~al.} 2006{\natexlab{a}},
  \apjl, 650, L41

\bibitem[{{Zucker} {et~al.}(2006{\natexlab{b}}){Zucker}, {Belokurov}, {Evans},
  {Wilkinson}, {Irwin}, {Sivarani}, {Hodgkin}, {Bramich}, {Irwin}, {Gilmore},
  {Willman}, {Vidrih}, {Fellhauer}, {Hewett}, {Beers}, {Bell}, {Grebel},
  {Schneider}, {Newberg}, {Wyse}, {Rockosi}, {Yanny}, {Lupton}, {Smith},
  {Barentine}, {Brewington}, {Brinkmann}, {Harvanek}, {Kleinman}, {Krzesinski},
  {Long}, {Nitta}, \& {Snedden}}]{2006ApJ...643L.103Z}
---. 2006{\natexlab{b}}, \apjl, 643, L103

\end{thebibliography}

\end{document}